\documentclass[traditabstract]{aa}

\usepackage{graphicx}
\usepackage{txfonts}
\usepackage{xcolor}
\usepackage{comment}
\usepackage{array}
\usepackage{natbib}
\bibpunct{(}{)}{;}{a}{}{,}
\usepackage{hyperref}
\hypersetup{colorlinks=true,
            citecolor=blue,
            linkcolor=blue}
\usepackage{color}

\newcolumntype{H}{>{\setbox0=\hbox\bgroup}c<{\egroup}@{}}

\def\be {\begin{equation}}
\def\ee {\end{equation}}
\def\mp {m}
\def\Rp {R}

\def\ms {M}
\def\nG {H}
\def\vG {\vec{\nG}}
\def\nL {L}
\def\vL {\vec{\nL}}
\def\vT {\vec{T}}
\def\Gc {{\cal G}}
\def\vk {\vec{k}}
\def\vp {\vec{p}}
\def\vq {\vec{q}}
\def\vs {\vec{s}}
\def\vg {\vec{g}}
\def\bk {b}
\def\om {\omega}
\def\At {{\cal K}}

\def\vw {\vec{\omega}}

\def\vr {\vec{r}}
\def\ur {\vec{\hat r}}

\def\ii {\mathrm{i}} 
\def\lv {\upsilon} 
\def \cT {x}
\def \Xtz {X_k^{-3,0}}
\def \Xtdp {X_k^{-3,2}}
\def \Xtdm {X_k^{-3,-2}}
\def \ke {k_{\rm e}}
\def \kf {k_{\rm f}}
\def \te {\tau_e}
\def \tv {\tau_v}
\def \nt {n \tau}
\def \sq {\sqrt{1-e^2}}
  
\def\figpath{}
\def \llabel#1{\label{#1}}
\def\bfx#1{#1}

\begin{document}

\title{Tidal excitation of the obliquity of Earth-like planets in the habitable zone of M-dwarf stars}
\titlerunning{Tidal excitation of the obliquity of Earth-like planets}

\authorrunning{E. F. S. Valente \& A. C. M. Correia }
\author{
Ema F. S. Valente\inst{1}
 \and
Alexandre C. M. Correia\inst{1,2}
}

\institute{
CFisUC, Departamento de F\'isica, Universidade de Coimbra, 3004-516 Coimbra, Portugal
\and 
IMCCE, Observatoire de Paris, PSL Universit\'e, 77 Av. Denfert-Rochereau, 75014 Paris, France
}

\date{\today; Received; accepted To be inserted later}

  \abstract{
Close-in planets undergo strong tidal interactions with the parent star that modify their spins and orbits.
In the two-body problem, the final stage for tidal evolution is the synchronisation of the rotation and orbital periods, and the alignment of the planet spin axis with the normal to the orbit (zero planet obliquity).
The orbital eccentricity is also damped to zero, but over a much longer timescale, that may exceed the lifetime of the system.
For non-zero eccentricities, the rotation rate can be trapped in spin--orbit resonances that delay the evolution towards the synchronous state.
Here we show that capture in some spin--orbit resonances may also excite the obliquity to high values rather than damp it to zero.
Depending on the system parameters, obliquities of $60^\circ-80^\circ$ can be maintained throughout the entire lifetime of the planet.
This unexpected behaviour is particularly important for Earth-like planets in the habitable zone of M-dwarf stars, as it may help to sustain temperate environments and thus more favourable conditions for life.
}


   \maketitle
%

\section{Introduction}
\label{Intro}

The discovery of a large number of exoplanets over recent decades shows that they are common around main sequence dwarf stars \citep[e.g.][]{Winn_2018}. 
The most abundant planets are the super-Earths, that is, planets with greater masses than that of the Earth and smaller than Neptune \citep[e.g.][]{Schlichting_2018}.
However, there exists an observational bias towards large masses, and so the population of Earth-mass planets is also expected to be large, as shown by formation studies \citep[e.g.][]{Schlecker_etal_2021}.
The fact that Earth-like planets may exist around other stars offers a great opportunity to explore the possibility of having life as we know it outside the Solar System.
        
A configuration like the Earth around the Sun is ideal when searching for habitable worlds.
However, \bfx{researchers have shown interest} in studying systems where the host stars are smaller than our Sun, the M-dwarfs.
These stars are the most abundant in the Milky Way, representing $\sim 70\%$ of all stars \citep{Bochanski_2010}, and the occurrence rate of planets is higher than around other main sequence stars \citep{Tuomi_etal_2019}.
Formation studies also show that temperate, Earth-sized planets are most frequent around early M-dwarfs \citep{Burn_etal_2021}.
Moreover, these stars are perfect for detecting low-mass planets via radial velocity and transit techniques \citep{Tuomi_2014, Shields_2016} because of their smaller mass and size.
Indeed, several planets in the habitable zone of M-dwarfs have already been discovered, some of which are promising candidates to harbour life \citep[e.g.][]{Anglada_Escude_2013, Anglada_Escude_etal_2016, Dittmann_2017, Gillon_2017, Bonfils_2018,Reiners_2018, Zechmeister_etal_2019}. 

Searching for habitable worlds around M-dwarfs has some advantages. 
Because of their lower luminosities, the region where planets can sustain liquid water, usually called the habitable zone \citep{Kasting_etal_1993}, is closer to the star.
In addition, these are the longest-living stars in the Universe, which provides enough \bfx{time for planetary and biological development} \citep{Shields_2016}. 
However, it also has some disadvantages. 
M-dwarfs stars are very active stars, especially in their earlier lifetime, which drives atmospheric escape \citep[e.g.][]{France_etal_2020}.
The only possibility to avoid this is the presence of a strong magnetic field on the planets \citep{Lammer_etal_2007}, or a late formation of secondary atmospheres \citep{Kite_Barnett_2020}.
Another concern is that, as the habitable zone is closer to the star, the planets undergo strong tidal effects, which modify their spins and orbits.

The spin (rotation and obliquity) and the orbit (semi-major axis and eccentricity) of a planet are very important for habitability, because they control the heat distribution on the surface of the planet, known as insolation \citep[e.g.][]{Milankovitch_1941, Ward_1974}. 
The stellar flux is inversely proportional to the square of the semi-major axis, and so this parameter is used to define the limits of the habitable zone.
High eccentricities also change the stellar flux on average, because of the different distances at the pericentre and appocentre \citep[e.g.][]{Dobrovolskis_2007}.
Therefore, \bfx{planets with eccentric orbits} in the habitable zone may not be able to sustain liquid water on the surface for the whole orbital period \citep{Bolmont_2016}.

The rotation rate of a planet affects the day--night cycle, but also the atmosphere through the Coriolis effect. 
For instance, for slow-rotating planets (like Venus) this effect is weak, which promotes the creation of thick stationary clouds in the substellar region, reflecting the solar rays \citep[e.g.][]{Kopparapu_2019}. 
On the other hand, for fast-rotating planets (like the Earth) this effect is stronger and \bfx{the clouds are in bands} on the tropics, which in turn reduces the albedo \citep[e.g.][]{Yang_2014}.

The obliquity controls the insolation distribution for a given latitude, and is therefore the main driver of the seasons.
High obliquities promote a more balanced distribution of the stellar flux on the planet surface, and thus extend the size of the habitable areas.
Indeed, for obliquities of around $50^\circ$ and $130^\circ$, the equatorial and polar regions receive identical amounts of insolation on average, while for obliquities close to $90^\circ$ the polar regions receive as \bfx{much} insolation as the equatorial regions when the obliquity is $0^\circ$ \citep[e.g.][]{Dobrovolskis_2021}.
Moreover, global climate model simulations show that high-obliquity planets are hotter than their low-obliquity counterparts because of ice--albedo feedbacks for cold climates, which extends the outer edge of the habitable zone \citep{Armstrong_2014, Colose_2019}.

The final state of tidal evolution corresponds to circular orbits, zero obliquity, and synchronous rotation \citep{Hut_1980, Adams_Bloch_2015}.
In this configuration, the rotation period is equal to the orbital period, implying that one side of the planet always faces the star and becomes extremely hot, while the other side becomes very cold.
Though not impossible, this extreme environment is not suitable for the development of life as we know it \citep{Wordsworth_2015,Turbet_2016}. 
However, the eccentricity of planets in the habitable zone of M-dwarfs is expected to evolve slowly, which allows the emergence of spin--orbit resonances that delay the evolution to the synchronous state \citep{Makarov_etal_2012, Correia_etal_2014}.
Moreover, for multi-planet systems, the eccentricity is excited by mutual perturbations and can keep a non-zero value \citep{Laskar_etal_2012, Barnes_2017}.
In multi-planet systems, the obliquity can also be locked in a `Cassini state', where it retains a non-zero value \citep{Colombo_1966}.
These equilibria occur more often for \bfx{low obliquities}, but they can also reach high values for systems with companions of large mass or high mutual inclination \citep{Correia_2015}. 
Resonant spin--orbit coupling can also trigger large chaotic variations of the obliquity \citep{Laskar_Robutel_1993, Su_Lai_2022b}.

Although \bfx{high obliquities} are possible in multi-body systems, for Earth-like close-in planets in the habitable zone of M-dwarfs, non-zero equilibrium states can be easily broken and evolve to a near-zero obliquity value \citep{Levrard_etal_2007, Millholland_Laughlin_2019, Su_Lai_2022a}.
In this paper we describe a new mechanism that can stabilise the obliquity of these planets at high values, which is valid in the two-body problem and does not require perturbations from other companions.

In Sect.~\ref{methods}, we present a general model to study the tidal evolution of close-in planets obtained in \citet{Correia_Valente_2022}.
In Sect.~\ref{obli}, we analyse the secular evolution of the rotation and obliquity in a very general framework, and show under which circumstances the obliquity is allowed to grow to high values.
In Sect.~\ref{apliRoss}, we apply our model to an Earth-mass planet in the habitable zone of an M-dwarf star, Ross~128\,b \citep{Bonfils_2018}, and confirm the predictions from our model.
Finally, \bfx{in the last section} we summarise and discuss our results.

\section{Model}
\label{methods}

We consider a system in a Keplerian orbit, composed of a star and a planet, with masses $\ms$ and $\mp$, respectively.
The orbital energy and angular momentum are respectively given by \citep[e.g.][]{Murray_Dermott_1999}
\be
E_{\rm orb} = - \frac{\Gc \ms \mp}{2 a} 
\ ,\quad \mathrm{and} \quad
\vG = \beta n a^2 \sq \, \vk 
\ , \label{210804a}
\ee
where
$\Gc$ is the gravitational constant,
$a$ is the semi-major axis, 
$e$ is the eccentricity,
$n=[\Gc (\ms + \mp) a^{-3}]^{1/2}$ is the mean motion,
$\beta = \ms \mp (\ms + \mp)^{-1}$ \bfx{is the reduced mass},
and $\vk$ is the unit vector along the direction of $\vG$, which is normal to the orbit.
The star is a point-mass object, while the planet is an extended body with radius $\Rp$, that can be deformed under the action of tides.
The planet rotates with angular velocity $\vw=\om \, \vs$, where $\vs$ is the unit vector along the direction of the spin axis. 
We assume that $\vs$ is also the axis of maximal inertia (gyroscopic approximation), and so the rotational angular momentum is simply given by
\be
\vL = C \vw \ , \llabel{210804b}
\ee
where $C = \xi \mp \Rp^2$ is the principal moment of inertia of the planet and $\xi$ an internal structure constant.

\subsection{Tidal torque and power}

Estimations of the spin and orbital evolution of the planet are based on a very general formulation of the tidal potential initiated by \citet{Darwin_1879a}.
Tidal effects arise from differential and inelastic deformations of the planet due to the gravitational perturbations from the star.
The distortion of the planet gives rise to a tidal potential \citep[e.g.][]{Lambeck_1980},
\be 
V (\vr) = - k_2 \frac{\Gc \ms}{\Rp} \left(\frac{\Rp}{r}\right)^3 \left(\frac{\Rp}{r_\star}\right)^3 P_2 ( \ur \cdot \ur_\star) \ , \llabel{210805a}
\ee
where $\vr$ is the distance measured from the planet centre of mass,
$r = ||\vr||$ is the norm, $\ur = \vr / r $ is the unit vector, $\vr_\star$ is the position of the star, and $P_2(x) = (3 x^2 -1)/2 $ is a Legendre polynomial.
$k_2$ is the second Love number for potential; it is a complex number, and depends on the frequency of the perturbation, $\sigma$.
We can decompose $k_2$ in its real and imaginary parts as
\be
 k_2 (\sigma) = a (\sigma) - \ii \, \bk(\sigma) 
\ . \llabel{210924d}
\ee
This partition is very useful when we write the equations of motion (see Sect.~\ref{doubav}), because the imaginary part characterises the viscous phase lag of the  material and is therefore directly related to the amount of energy dissipated by tides.
In general, the deformation lags behind the perturbation and therefore the imaginary part is always negative, hence the minus sign in expression \eqref{210924d}.

The tidal potential (Eq.\,\eqref{210805a}) creates a differential gravitational field around the planet given by
\be
\vg (\vr) = - \nabla_{\vr} V (\vr) \ .
\ee
The star itself interacts with this field; with mass $\ms$ and located at $\vr=\vr_\star$,  it exerts a torque on the planet that modifies its spin and orbit.
In an inertial frame we have 
\be
\dot \vG = \vT = \ms \, \vr_\star \times \vg  (\vr_\star)
\ , \llabel{150626a}
\ee
and, owing to the conservation of the total angular momentum,
\be
\dot \vL = - \dot \vG = - \vT
\ . \llabel{210805b}
\ee
The evolution of the orbital energy (power) is given by
\be
\dot E_{\rm orb} = \ms \, \dot \vr_\star \cdot \vg (\vr_\star)
\ . \llabel{211110b}
\ee

\subsection{Secular equations of motion}
\llabel{doubav}       

In general, tidal effects slowly modify the spin and the orbit of the planet, in a timescale much longer than the orbital and precession periods of the system.
We can then average the torque (Eq.\,\eqref{150626a}) and the power (Eq.\,\eqref{211110b}) over the mean anomaly and the argument of the pericentre, and obtain the equations of motion for the secular evolution of the system.

We let ($\vp,\vq,\vs$) be a cartesian reference frame, such that
\be
\vp = \frac{ \vk \times \vs }{\sin \theta} \ , 
\quad \vq = \frac{\vk - \cos \theta \, \vs}{\sin \theta}  \ , 
\quad \cos \theta = \vk \cdot \vs \ ,
\ee
where $\vp$ is aligned with the line of nodes between the equator of the planet and the orbital plane, and $\theta$ is the angle between these two planes, also known as the obliquity.
Following \citet{Correia_Valente_2022},  for the average torque, we therefore have
\be
\big\langle \vT \big\rangle =  T_p \sin \theta \, \vp + T_q \sin \theta \, \vq + T_s \, \vs 
\ , \llabel{220405a}
\ee
and for the average power
\be
\llabel{220405b}
\big\langle \dot E_{\rm orb} \big\rangle = n \, T_E \ ,
\ee
with\footnote{We do not provide the expression for $T_p$, because in our case this component changes neither the norm of the angular momentum vectors nor the angle between them (Sect.~\ref{orbspinevol}). This is also why we do not take into account the effect of the rotational deformation, as it only contributes to the torque with a component along $\vp$ \citep[e.g.][]{Correia_2006}.} 
\be
\begin{split}
\llabel{211110t1}
T_q = \frac{3 \At}{32} \sum_{k=-\infty}^{+\infty} & \Bigg\{ 3 \, \bk ( -kn ) \left(1 - \cT^2 \right) \bigg[ \left( \Xtdp \right)^2 - \left( \Xtdm \right)^2 \bigg]  \\ 
   &  + 2 \, \bk ( \om-kn ) \bigg[  \left( 1+ \cT \right)^2 \left( 2-\cT \right) \left( \Xtdp \right)^2  \\ 
   &  - 4 \cT^3 \left( \Xtz \right)^2  - \left( 1- \cT \right)^2 \left( 2+\cT \right) \left( \Xtdm  \right)^2  \bigg]  \\
   & + \bk \left( 2\om-kn \right) \bigg[ - 4 \cT \left( 1- \cT^2  \right) \left( \Xtz \right)^2 \\
   &  + \left( 1+ \cT \right)^3 \left( \Xtdp \right)^2 - \left( 1 - \cT \right)^3 \left( \Xtdm \right)^2  \bigg] \Bigg\}  \ ,
\end{split}
\ee
\be
\llabel{211110t5}
\begin{split}
T_s = \frac{3 \At}{32} \sum_{k=-\infty}^{+\infty} & \Bigg\{ 2 \, \bk ( \om-kn ) \left( 1 - \cT^2 \right) \bigg[ 4 \cT^2 \left( \Xtz  \right)^2   \\
   & + \left( 1 + \cT \right)^2 \left( \Xtdp \right)^2 + \left( 1 -\cT \right)^2 \left( \Xtdm  \right)^2  \bigg] \\
   & + \bk ( 2\om-kn ) \bigg[ 4 \left( 1 - \cT^2 \right)^2 \left( \Xtz \right)^2  \\
   & + \left( 1 + \cT  \right)^4 \left( \Xtdp  \right)^2 + \left( 1 - \cT  \right)^4 \left( \Xtdm  \right)^2  \bigg] \Bigg\}   \ ,
\end{split}
\ee
\be
\llabel{211110t4}
\begin{split}
T_E = \frac{\At}{64}  \sum_{k=-\infty}^{+\infty} & k \, \Bigg\{ \bk ( -kn )  \bigg[
   4 \left(1-3 \cT^2\right)^2 \left(\Xtz\right)^2 \\
& + 9 \left(1-\cT^2\right)^2 \left(\left(\Xtdm\right)^2 + \left(\Xtdp\right)^2\right) \bigg]  \\
& + 12 \, \bk ( \om-kn ) \left(1-\cT^2\right) \bigg[  4 \cT^2  \left(\Xtz\right)^2 \\
& +  \left(1-\cT\right)^2 \left(\Xtdm\right)^2 + \left(1+\cT\right)^2 \left(\Xtdp\right)^2  \bigg]  \\
& + 3 \, \bk ( 2\om-kn )  \bigg[ 4 \left(1-\cT^2\right)^2 \left(\Xtz\right)^2 \\
& + \left(1-\cT\right)^4 \left(\Xtdm\right)^2 + \left(1+\cT\right)^4 \left(\Xtdp\right)^2 \bigg] \Bigg\} \ ,
\end{split}
\ee
where $ \At = \Gc \ms^2 R^5 a^{-6}$, $\cT=\cos \theta$, and $X_k^{-3,m} (e)$ are the Hansen coefficients, which depend only on the eccentricity (see Table~\ref{TabX}).
Throughout this work, we consider only terms with $|k| \le 20$, because for the maximal adopted eccentricity ($e=0.2$), we get an error smaller than $e^{20} \approx 10^{-14}$, which corresponds to the computer precision that we have in our simulations.

\begin{table}
\caption{Hansen coefficients $X_k^{-3,0} (e) $ and $X_k^{-3,2} (e) $  up to $ e^6 $. \label{TabX} } 
\begin{center}
\begin{tabular}{|r|c|c| } \hline 
$k$ & $X_k^{-3,0} (e) $ & $X_k^{-3,2} (e) $ \\ \hline
$-6$ & $\frac{3167}{320} e^6$ & $-$ \\
$-5$ & $\frac{1773}{256} e^5$ & $-$ \\
$-4$ & $\frac{77}{16} e^4 + \frac{129}{160} e^6$ & $\frac{4}{45} e^6$ \\
$-3$ & $\frac{53}{16} e^3 + \frac{393}{256} e^5$ & $\frac{81}{1280} e^5$ \\
$-2$ & $\frac{9}{4} e^2 + \frac{7}{4} e^4 + \frac{141}{64} e^6$ & $\frac{1}{24} e^4 + \frac{7}{240} e^6$ \\
$-1$ & $\frac{3}{2} e + \frac{27}{16} e^3 + \frac{261}{128} e^5$ & $\frac{1}{48} e^3 + \frac{11}{768} e^5$ \\
$0$ & $1 + \frac{3}{2} e^2 + \frac{15}{8} e^4 + \frac{35}{16} e^6$ & $-$ \\
$1$ & $\frac{3}{2} e + \frac{27}{16} e^3 + \frac{261}{128} e^5$ & $- \frac{1}{2} e + \frac{1}{16} e^3 - \frac{5}{384} e^5$ \\
$2$ & $\frac{9}{4} e^2 + \frac{7}{4} e^4 + \frac{141}{64} e^6$ & $1 - \frac{5}{2} e^2 + \frac{13}{16} e^4 - \frac{35}{288} e^6$ \\
$3$ & $\frac{53}{16} e^3 + \frac{393}{256} e^5$ & $\frac{7}{2} e - \frac{123}{16} e^3 + \frac{489}{128} e^5$ \\
$4$ & $\frac{77}{16} e^4 + \frac{129}{160} e^6$ & $\frac{17}{2} e^2 - \frac{115}{6} e^4 + \frac{601}{48} e^6$ \\
$5$ & $\frac{1773}{256} e^5$ & $\frac{845}{48} e^3 - \frac{32525}{768} e^5$ \\
$6$ & $\frac{3167}{320} e^6$ & $\frac{533}{16} e^4 - \frac{13827}{160} e^6$ \\
$7$ & $-$ & $\frac{228347}{3840} e^5$ \\
$8$ & $-$ & $\frac{73369}{720} e^6$ \\ \hline
\end{tabular} 
\end{center}
$X_k^{-3,-2} (e) = X_{-k}^{-3,2} (e)$. The exact expression of these coefficients is given by $  X_k^{-3,m} (e)  = \pi^{-1} \int_0^\pi \left( a/r \right)^{3} \exp(\ii m \lv) \exp(-\ii k M) \, d M $. 
\end{table}

\subsection{Orbital and spin evolution}
\llabel{orbspinevol}

The secular evolution of the orbital elements is obtained from the orbital energy and angular momentum (Eq.\,\eqref{210804a}) as
\be
\frac{\dot a}{a}  = \frac{2 a}{\Gc \ms \mp} \, \big\langle \dot E_{\rm orb} \big\rangle = \frac{2 \sq}{\nG} \, T_E
 \ , \llabel{220406c}
\ee
\be
\begin{split}
\dot e & = \frac{(1-e^2)}{2 e} \, \frac{\dot a}{a} - \frac{\sq}{\beta n a^2 e} \, \big\langle \vT \big\rangle \cdot \vk \\
& = \frac{1-e^2}{\nG e} \left( T_E \sq - T_q \sin^2 \theta - T_s \cos\theta  \right) 
 \ , \llabel{220406d}
\end{split}
\ee
with $\nG =||\vG||$.
Similarly, the secular evolution of the rotation is obtained from the rotational angular momentum (Eq.\,\eqref{210804b}) as
\be
\frac{\dot \om}{\om} = - \frac{1}{C \om}\, \big\langle \vT \big\rangle  \cdot \vs = - \frac{T_s}{L}
\ , \llabel{220407a}
\ee
with $\nL=||\vL||$, 
while the secular evolution of the obliquity is obtained from $\cos \theta = \vk \cdot \vs = \vG \cdot \vL /(\nG \nL)$ as
\be
\begin{split}
\dot \theta & = \left[ \left( \frac{1}{\nL} + \frac{\cos\theta}{\nG} \right) \big\langle \vT \big\rangle \cdot \vk - \left( \frac{1}{\nG} + \frac{\cos\theta}{L} \right) \big\langle \vT \big\rangle \cdot \vs \right] \csc\theta \\
& =  \left[ \left( \frac{1}{\nL} + \frac{\cos\theta}{\nG}  \right) T_q 
- \frac{T_s}{\nG}  \right] \sin\theta
\ . \llabel{220407b}
\end{split}
\ee
In general, we have $\nL \ll \nG$ (Eq.\,\eqref{220414b}), and therefore the previous expression for the obliquity can be simplified as
\be
\dot \theta \approx  \frac{T_q}{\nL} \sin\theta
\ . \llabel{220407c}
\ee

\subsection{Maxwell rheology}
\llabel{tidalmodels}

The tidal deformation is characterised by the Love number (Eq.\,\eqref{210924d}), which depends on \bfx{the internal structure of the planet} and is therefore subject to large uncertainties.
For this reason, in order to compute $ k_2 (\sigma)$ we need to adopt some rheological model for the deformation.
A large variety of models exist, but the most commonly used are the constant$-Q$ \citep[e.g.][]{Munk_MacDonald_1960}, the linear model \citep[e.g.][]{Mignard_1979}, 
the Maxwell model \citep[e.g.][]{Correia_etal_2014},
and the Andrade model \citep[e.g.][]{Efroimsky_2012b}.
Some models appear to be better adapted to certain situations, but there is no model that is globally employed.
However, viscoelastic rheologies are usually accepted as more realistic for rocky planets, because they are able to simultaneously reproduce the short-time and the long-time responses \citep[e.g.][]{Remus_etal_2012b}.
A review of the main viscoelastic models can be found in \citet{Renaud_Henning_2018}.

We adopt here a Maxwell viscoelastic model, which is particularly well suited to reproducing the long-term deformation of the planets.
A material is called a Maxwell solid when it responds to stresses like a massless, damped harmonic oscillator \citep[e.g.][]{Turcotte_Schubert_2002}.
It is characterised by a rigidity $\mu$ (or shear modulus), and by an effective viscosity $\eta$. 
Over short timescales,
the material behaves like an elastic solid, but over long periods of time it flows like a fluid.
The Love number for the Maxwell model is given by \citep[e.g.][]{Darwin_1908}
\be
k_2 (\sigma) = \kf \, \frac{1 + \ii \sigma \te}{1 + \ii \sigma \tau}
\ , \quad \mathrm{with} \quad
\tau = \te + \tv
\ , \llabel{210929a}
\ee
where
$\tau$, $\tv$, and $\te = \eta / \mu$ are the total, the viscous, and the Maxwell relaxation times, respectively. 
Here, $\kf = k_2(0)$ is the fluid Love number, which corresponds to the maximal deformation that is attained for long-term perturbations.
Similarly, we can also define an elastic Love number, $\ke = k_2 (\infty) = \kf \, \te / \tau$, which corresponds to the deformation for short-time perturbations.
Both $\kf$ and $\ke$ depend only on the internal structure of the planet, and can usually be measured for Solar System rocky planets from the rotational and tidal \bfx{deformations}, respectively \citep[e.g.][]{Correia_Rodriguez_2013}.
Therefore (Eqs.\,\eqref{210924d} and \eqref{210929a}),
\be
\bk(\sigma) 
= \kf \, \frac{\sigma \tv}{1 + (\sigma \tau)^2} 
= \left(\kf - \ke \right) \, \frac{\sigma \tau }{1 + (\sigma \tau)^2} 
\ . \llabel{210929z}
\ee
For the Earth, we have $\ke = 0.295$ and $\kf = 0.933$ \citep{Yoder_1995cnt}.
The total relaxation time can be obtained as $\tau = \te \, \kf / \ke \approx 3 \eta / \mu$.
The rigidity of the Earth's mantle is relatively well constrained, $\mu \approx 80$~GPa \citep{Karato_Wu_1993}.
However, the average effective viscosity is much more uncertain, $\eta \sim 10^{21}$~Pa\,s \citep{Karato_Wu_1993}, which gives $\te \sim 400$~yr.
Indeed, the viscosity may vary by many orders of magnitude, depending on the depth and temperature \citep[e.g.][]{Kirby_Kronenberg_1987}.
In the case of the Earth, the surface post-glacial rebound due to the last glaciation about $10^4$~yr ago is still ongoing, suggesting that the Earth's mantle relaxation time is $\te \approx 4400$~yr \citep{Turcotte_Schubert_2002}.
On the other hand, for slightly warmer planets we may have lower values, such as $\te \sim 50$~yr \citep[e.g.][]{Makarov_etal_2012}.


\section{Spin dynamics}
\label{obli}

In this section, we analyse the dynamics of the system provided by the secular equations (Sect.~\ref{orbspinevol}). 
The main goal is to describe the possible equilibria and evolutionary paths for different tidal regimes (different $\tau$ values). 

Regardless of the tidal model that we adopt, the final state of tidal evolution corresponds to circular orbits, zero obliquity, and synchronous rotation \citep{Hut_1980, Adams_Bloch_2015}.
However, the evolution timescales for the orbit and spin are quite different.
Indeed, from expressions \eqref{220406c}$-$\eqref{220407c}, we have
\be
\frac{\dot a}{a} \sim \dot e \sim \frac{\At}{\nG} 
\quad \ll \quad
\frac{\dot \om}{\om} \sim \dot \theta \sim \frac{\At}{\nL} \ , 
\llabel{220414a}
\ee
since (Eqs.\,\eqref{210804a} and \eqref{210804b})
\be
\frac{\nL}{\nG} = \frac{C \om}{\beta n a^2 \sq} \approx \frac{\om}{n} \left(\frac{R}{a}\right)^2 \ll 1
\ . \llabel{220414b}
\ee
As a result, the spin evolves much faster than the orbit and is allowed to reach metastable or pseudo-equilibrium states as long as the eccentricity is not completely zero.
Therefore, in this section we assume a constant value for the semi-major axis and eccentricity, and solely study the spin evolution.

We limit our analysis to positive rotation rates ($\omega>0$), but we can easily extend it to negative rotations with the symmetry $(\omega, \theta) \Leftrightarrow (-\omega, \pi-\theta)$. 
These two points do not correspond to the same physical state, but are equivalent from a dynamical point of view \citep{Correia_Laskar_2001}.

We also note that the secular evolution of the spin (Eqs.\,\eqref{220407a} and \eqref{220407c}) depends only on four parameters:
the rotation ratio, $\om / n$, 
the obliquity, $\theta$,
the eccentricity, $e$,
and the relative relaxation time, $\nt$ (Eq.\,\eqref{210929z}).
The amplitude $\At/L$ can change the evolution rate,
but does not modify the spin dynamics.
        
\subsection{Equilibrium rotation}
\llabel{eqrot}

The final evolution of the spin has zero obliquity ($\theta=0$), which is also an equilibrium point (Eq.\,\eqref{220407c}).
Therefore, for simplicity previous studies usually fix $\theta=0$, that is, they adopt a planar model \citep[e.g.][]{Makarov_etal_2012, Ferraz-Mello_2013, Correia_etal_2014}.
\bfx{This is equivalent to setting $x=1$} in expression \eqref{211110t5}, and thus expression \eqref{220407a} simplifies as
\be
\frac{\dot \om}{\om} 
= - \frac{3\At}{2 \nL}  \sum_{k=-\infty}^{+\infty}  \bk(2 \om - k n) \left(\Xtdp (e)\right)^2 
\ . \llabel{211103b}
\ee

\begin{figure*}
\centering
\includegraphics[width=\textwidth]{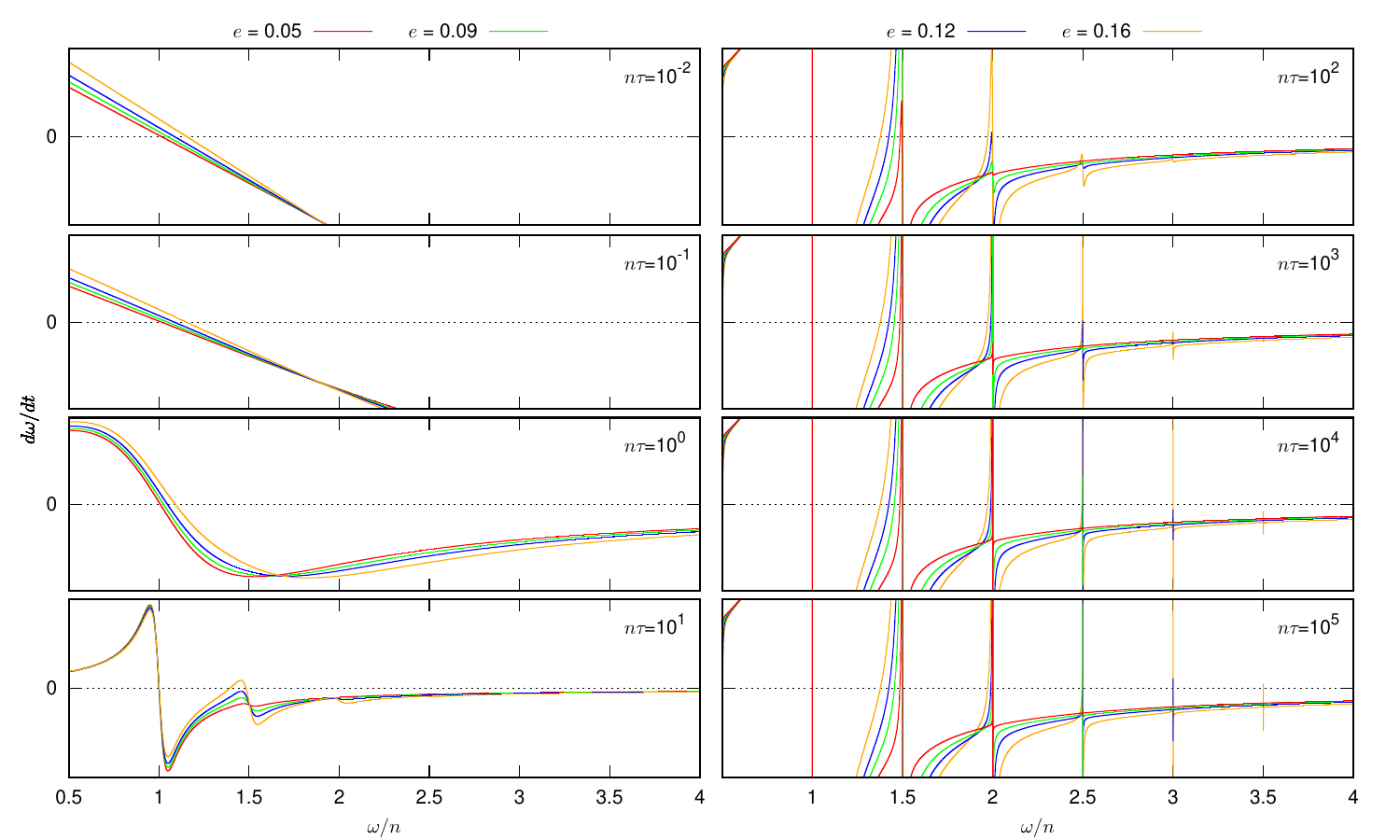}
\caption{\bfx{Variation of $d \omega / dt$ with $\omega / n$} for different values of $\nt$ and four different eccentricities: $e=0.05$ (red), $e=0.09$ (green), $e=0.12$ (blue), and $e=0.16$ (orange). We adopt $\theta = 0$, and the Hansen coefficients $\Xtdp(e)$ are truncated at $|k| \leq 20$. \label{FigI1}}
\end{figure*}

In Fig.~\ref{FigI1}, we plot \bfx{the above expression} as a function of $\om/n$ for different values of the eccentricity and $n\tau$.
The interception with \bfx{the horizontal dotted line $\dot \om = 0$} gives all the equilibria for the rotation rate. 
Negative slopes correspond to stable equilibria, where the rotation can remain locked for a given value of eccentricity.
We adopt a maximal eccentricity $e=0.16$, because in this study we are only interested in close-in planets with small eccentricities.
Similar figures for higher eccentricity values can be found in \citet{Correia_etal_2014} and \citet{Ferraz-Mello_2015b}.

For $\nt \lesssim 1$, we observe that a single stable equilibrium point exists, close to synchronous rotation.
Indeed, for small tidal frequencies $(\nt \ll 1)$, the Maxwell model behaves as the linear model, and expression \eqref{211103b} becomes \citep[e.g.][]{Correia_2009}
\be
\frac{\dot \om}{\om} 
\approx - \frac{3\At}{\nL} \left(\kf - \ke \right) \nt  \left(f_1(e) \frac{\om}{n} - f_2(e) \right) 
\ , \llabel{220419e}
\ee
with 
$f_1(e) \approx 1+ 15 e^2 /2$, and 
$f_2(e) \approx 1 + 27 e^2/2$.
The exact equilibrium, obtained when $\dot \om = 0$, is then given by 
\be
\frac{\om}{n} 
= \frac{f_2(e)}{f_1(e)}
\approx 1 + 6 e^2 
\ , \llabel{090520a}
\ee
which is also known as the pseudo-equilibrium rotation.

\begin{figure}
\centering
\includegraphics[width=\columnwidth]{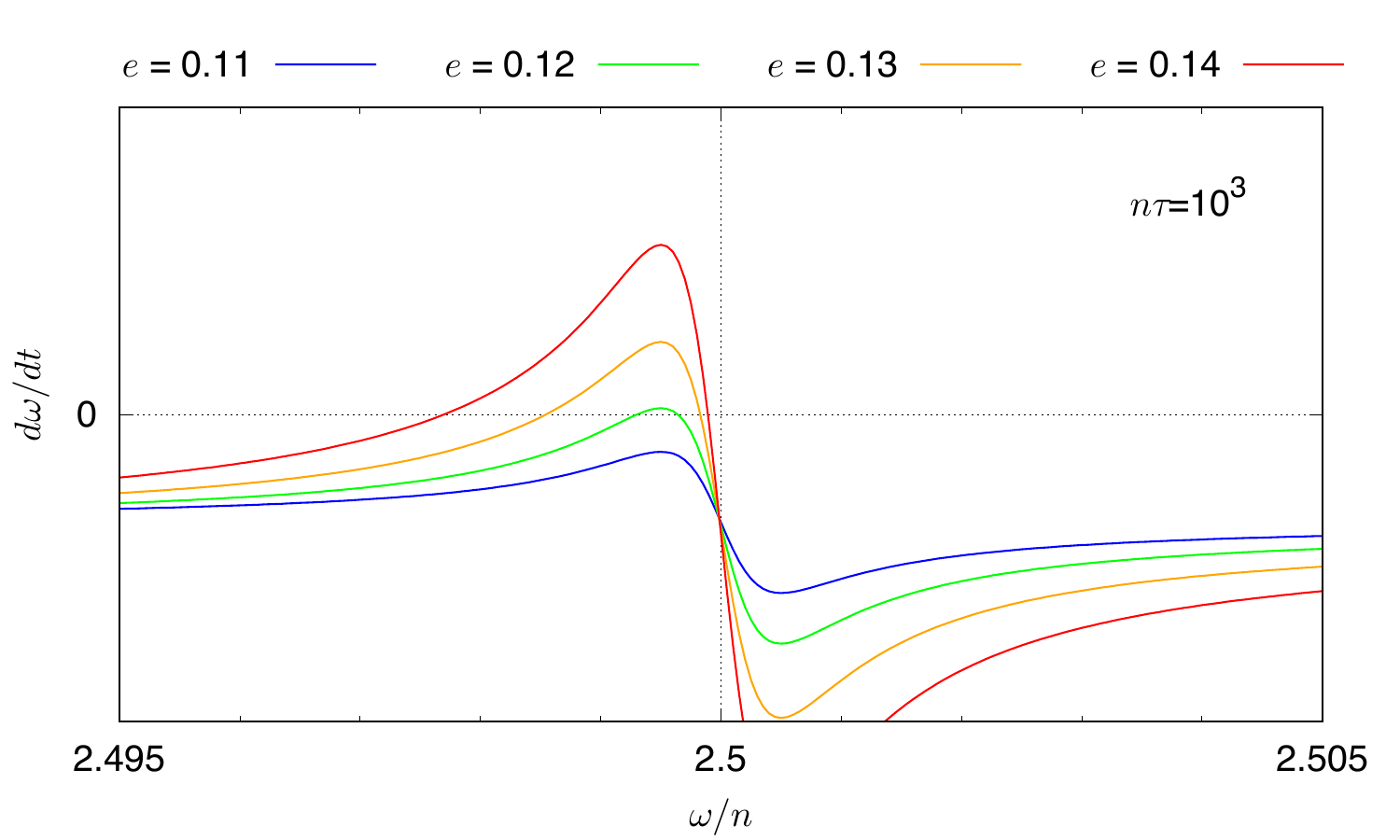}
\caption{\bfx{Variation of $d \omega / dt$ with $\omega / n$} around the $5/2$~resonance for $\nt=10^3$ and four different eccentricities: $e=0.11$ (blue), $e=0.12$ (green), $e=0.13$ (orange), and $e=0.14$ (red). We adopt $\theta = 0$, and the Hansen coefficients $\Xtdp(e)$ are truncated at $|k| \leq 20$. \label{zoomdw}}
\end{figure}

For $\nt \gg 1$, we observe that additional stable equilibria appear.
They cluster around values $\om/n \approx k/2$, because expression \eqref{211103b} has extrema at these values: 
\be
\frac{\dot \om}{\om} 
\approx
 - \frac{3\At}{4 \nL} \frac{ \left(\kf - \ke \right)}{\nt}  \sum_{k=-\infty}^{+\infty} \, \frac{\left(\Xtdp (e)\right)^2}{\om/n - k/2}  
\ . \llabel{220418a}
\ee
For all $k/2$ values, $\dot \om$ undergoes some oscillation, and its amplitude depends on $\Xtdp(e)$.
The rotation is locked whenever the oscillation amplitude surpasses the line $\dot\om = 0$.
These new states occur at almost semi-integer values ($k \in \mathbb{Z}$), and they are usually called spin--orbit resonances \citep[e.g.][]{Goldreich_Peale_1966, Correia_Delisle_2019}.
The number of these equilibria increases with $\nt$, but also with the eccentricity value, because the amplitude of the Hansen coefficient $\Xtdp(e)$ increases with $e$ (Table~\ref{TabX}).
We note that, except for the synchronous resonance ($k=2$), for the remaining \bfx{stable} equilibria with $k>2$, we have $\dot \om = 0$ for a rotation rate that is slightly smaller than the exact resonant value, that is, for $\om = k n/2 - |\delta \om|$, with $|\delta \om / n| \ll 1$.
Indeed, as the oscillation in the torque introduced by the synchronous resonance has the largest amplitude (see Table~\ref{TabX}), the average of the oscillations in the resonances with $k>2$ is shifted to a negative value. 
In Fig.~\ref{zoomdw}, we show an example for the $5/2$~resonance with different eccentricities.
For $e = 0.11$, the torque is always negative, and therefore capture cannot occur in this resonance.
For $e \ge 0.12$, the torque surpasses the line $\dot \om = 0$, and the rotation can stabilise at the value with negative slope, which occurs for $\om/n = 5/2 - |\delta \om/n|$.

Immediately after their formation, the planets are supposed to rotate fast \citep[e.g.][]{Kokubo_Ida_2007}, that is, $\om/n \gg 1$.
Therefore, we can write $\bk(2\om-kn) \approx \bk(2\om)$ and expression \eqref{211103b} simplifies as
\be
\frac{\dot \om}{\om} = - \frac{3\At}{2 \nL} \, \bk(2 \om) f_1(e)
\ . \llabel{220419h}
\ee
As a consequence, we always have $\dot \om < 0$, which can also be seen in Fig.~\ref{FigI1} for $\om/n = 4$.
We thus approach the equilibria while decreasing from fast rotations (right hand side in Fig.~\ref{FigI1}), and higher order spin--orbit resonances are encountered first.

In Fig.~\ref{FigI1}, we observe that for the same $\nt$-value, stabilisation can occur at different spin--orbit resonances, depending on the eccentricity.
We are assuming here that the eccentricity is constant, but in reality it also slowly evolves (Eq.\,\eqref{220414a}).
As the eccentricity decreases, the higher order spin--orbit equilibria become unstable, and the rotation is sequentially captured in lower order resonances, until it reaches its final possibility for synchronous rotation (see Sect.~\ref{apliRoss}).
Therefore, the asynchronous equilibria are only metastable; however, depending on the system parameters, they can persist for a time longer than the age of the system.

\begin{figure*}
\centering
\includegraphics[width=\textwidth]{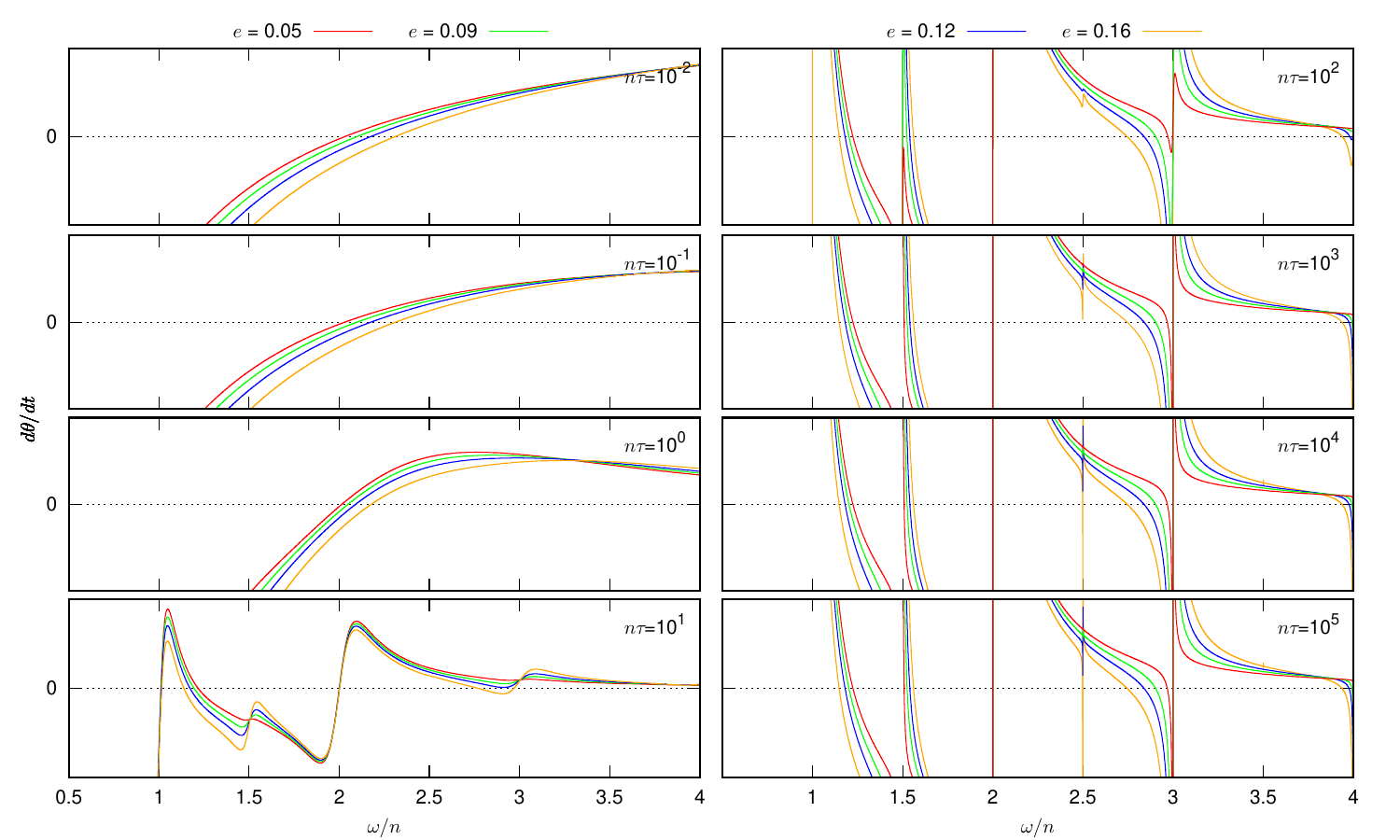}
\caption{\bfx{Variation of $d \theta / dt$ with $\omega / n$} for different values of $\nt$ and four different eccentricities: $e=0.05$ (red), $e=0.09$ (green), $e=0.12$ (blue), and $e=0.16$ (orange). We adopt $\theta \approx 0$, and the Hansen coefficients $X_k^{-3,m} (e)$ are truncated at $|k| \leq 20$. \label{FigI2}}
\end{figure*}

The general expression for the equilibria of the rotation rate ($T_s=0$) depends on the obliquity value (Eq.\,\eqref{211110t5}).
For non-zero obliquities, the possible equilibria for the rotation rate remain at the semi-integer values $\om/n = k/2$, because they correspond to the zeros of the $\bk(\sigma)$ functions.
However, as for the eccentricity, non-zero obliquities can also modify the amplitude of the torque and therefore lead to a different distribution of the available states for a given initial condition.

\subsection{Equilibrium obliquity}
\llabel{oblexct}

Although $\theta=0$ is always an equilibrium point (Eq.\,\eqref{220407c}), it is not necessarily stable.
For $T_q > 0$ (Eq.\,\eqref{211110t1}), this point is unstable and the obliquity is allowed to grow.
Therefore, in this section we linearise expression \eqref{220407c} for $\theta = \delta \theta \ll 1$ in order to determine when the planar approximation is no longer valid. 
We get
\be
\begin{split}
\llabel{220419a}
\dot \theta \approx  \frac{3 \At}{4 \nL} \sum_{k=-\infty}^{+\infty} & \Bigg\{ \bk ( \om-kn ) \, \bigg[ \left( \Xtdp (e) \right)^2  - \left( \Xtz (e) \right)^2 \bigg] \\
   & + \bk ( 2\om-kn ) \left( \Xtdp (e) \right)^2  \Bigg\} \, \delta \theta \ .
\end{split}
\ee
In Fig.~\ref{FigI2}, we plot the previous expression as a function of $\om/n$ for different values of the eccentricity and $n\tau$.
For $\dot \theta / \delta \theta < 0$, we have stable equilibrium at $\theta=0$, while for $\dot \theta / \delta \theta > 0$, this point is unstable.
For initial fast rotations ($\om/n \gg 1$), we can simplify expression \eqref{220419a} as \citep{Correia_Valente_2022}
\be
\dot \theta \approx  \frac{3 \At}{4 \nL} \, \bk ( 2\om ) f_1(e) \, \delta \theta 
\ . \llabel{220419i}
\ee
We conclude that we always have $\dot \theta / \delta \theta > 0$, 
which can also be seen in Fig.~\ref{FigI2} for $\om/n = 4$.
Therefore, $\theta=0$ is an unstable equilibrium point, and the obliquity increases to higher values.
However, as the rotation slows down owing to tides, we may have $\dot \theta / \delta \theta = 0$ for one or more $\om/n$ values, which represents a transition in the stability of the point $\theta=0$.

For $\nt \lesssim 1$, we observe that a single transition point exists, close to $\om/n = 2$.
In the linear model approximation $(\nt \ll 1)$, expression \eqref{220419a} becomes \citep[e.g.][]{Correia_2009}
\be
\dot \theta 
\approx  \frac{3 \At}{\nL} \left(\kf - \ke \right) \nt  \left( f_1(e) \frac{\om}{2 n}  - f_2(e) \right) \, \delta \theta 
\ , \llabel{220419f}
\ee
which gives for the exact position of the transition point,
\be
\frac{\om}{n} = 2 \, \frac{f_2(e)}{f_1(e)}
\approx 2 \, (1 + 6 e^2) 
\ . \llabel{220419c}
\ee
This value is two times larger than the equilibrium rotation (Eq.\,\eqref{090520a}).
As a result, although $\theta=0$ is initially unstable and the obliquity allowed to grow, it becomes stable as the rotation approaches its equilibrium value.

For $\nt \gg 1$, we observe that multiple transition points are present, introduced by an oscillation of the torque around the spin--orbit resonances $\om/n = k/2$, corresponding to the zeros of the $\bk(\sigma)$ functions.
As a result, the equilibrium at $\theta=0$ toggles between unstable and stable, and the transition points almost coincide with the spin--orbit resonances (Fig.~\ref{FigI1}).
\bfx{In general,} the torque on the obliquity (Eq.\,\eqref{220419a}) is negative for $\om/n < k/2$ and positive for $\om/n > k/2$ (Fig.~\ref{FigI2}).
Therefore, \bfx{for spin--orbit resonances, the obliquity} $\theta=0$ is usually stable, because for $k>2$, the equilibrium rotation occurs for $\om/n = k/2 - |\delta \om/n| < k/2 $ (Fig.~\ref{zoomdw}).
Nevertheless, there are a few exceptions, namely whenever the amplitude of the oscillation is not large enough to descend below the \bfx{horizontal line with} $\dot \theta = 0$.
In those cases, the torque is always positive, and $\theta=0$ remains unstable.
In Fig.~\ref{FigI2}, we observe that this is always the case for $\om/n = 7/2$, and it also occurs for $\om/n = 5/2$ for some eccentricities and $\nt$-values.
We therefore expect that the obliquity grows when the rotation rate is captured in these two specific spin--orbit resonances.

\begin{figure}
\centering
\includegraphics[width=\columnwidth]{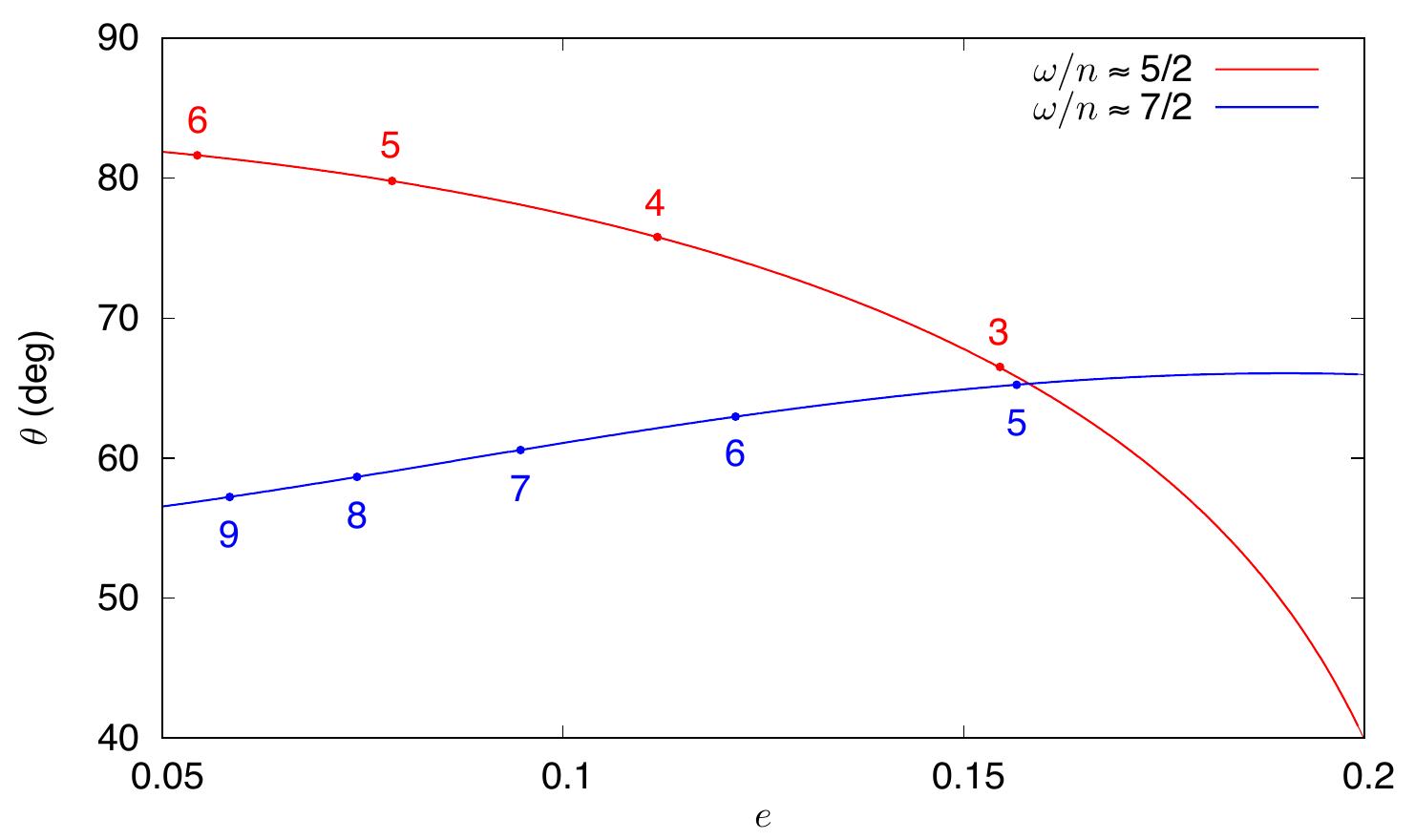}
\caption{High-obliquity stable equilibria as a function of eccentricity. The dots mark the critical eccentricity for $\nt=10^\alpha$, where the $\alpha$-value is given next to the dot. 
These states are found numerically by determining the pair ($\om/n, \theta$) that simultaneously satisfy $T_q=T_s=0$ (Eqs.\,\eqref{211110t1} and \eqref{211110t5}). The Hansen coefficients $X_k^{-3,m} (e)$ are truncated at $|k| \leq 20$. \label{eqobecc}}
\end{figure}

When the obliquity increases, expression \eqref{220419a} and Fig.~\ref{FigI2} are no longer valid.
$\theta = \pi$ is also an equilibrium point, because $\sin \pi = 0$ (Eq.\,\eqref{220407c}), but additional equilibria at intermediate obliquities may exist.
As the expressions of $T_q$ (Eq.\,\eqref{211110t1}) and $T_s$ (Eq.\,\eqref{211110t5}) are very sensitive to the value of $\om/n$, 
the additional equilibria for a given eccentricity and $\nt$-value can only be found numerically by determining the pair ($\om/n, \theta$) that simultaneously satisfy $T_q=T_s=0$.
In Fig.~\ref{eqobecc}, we compute the high-obliquity stable equilibria as a function of the eccentricity for different $\nt$-values.
We find that for $\om/n \approx 5/2$ and $\om/n \approx 7/2$, there exist high-obliquity states.
These states do not directly depend on the value of $\nt$, but they become unstable below a critical eccentricity, which depends on $\nt$. 
We observe that as we increase the $\nt$-value, the critical eccentricity becomes smaller.
For $\nt=10^5$, the obliquity can therefore reach nearly $80^\circ$ in the $5/2$~resonance and $65^\circ$ in the $7/2$~resonance.
These interesting high-obliquity stable equibria were previously unknown, and in principle can be attained for some evolutionary paths of the spin.

\subsection{Global spin evolution}
\llabel{globalspin}

The rotation rate and the obliquity evolution cannot be dissociated as they progress at a similar pace (Eq.\,\eqref{220414a}).
Therefore, the exact evolution of the spin can only be obtained by simultaneously integrating equations \eqref{220407a} and \eqref{220407c}.
This allows us to confirm the analytic predictions from Sects.~\ref{eqrot} and~\ref{oblexct}, in particular, the existence of non-zero obliquity equilibria.

\begin{figure*}
\centering
\includegraphics[width=\textwidth,height=0.93\textheight]{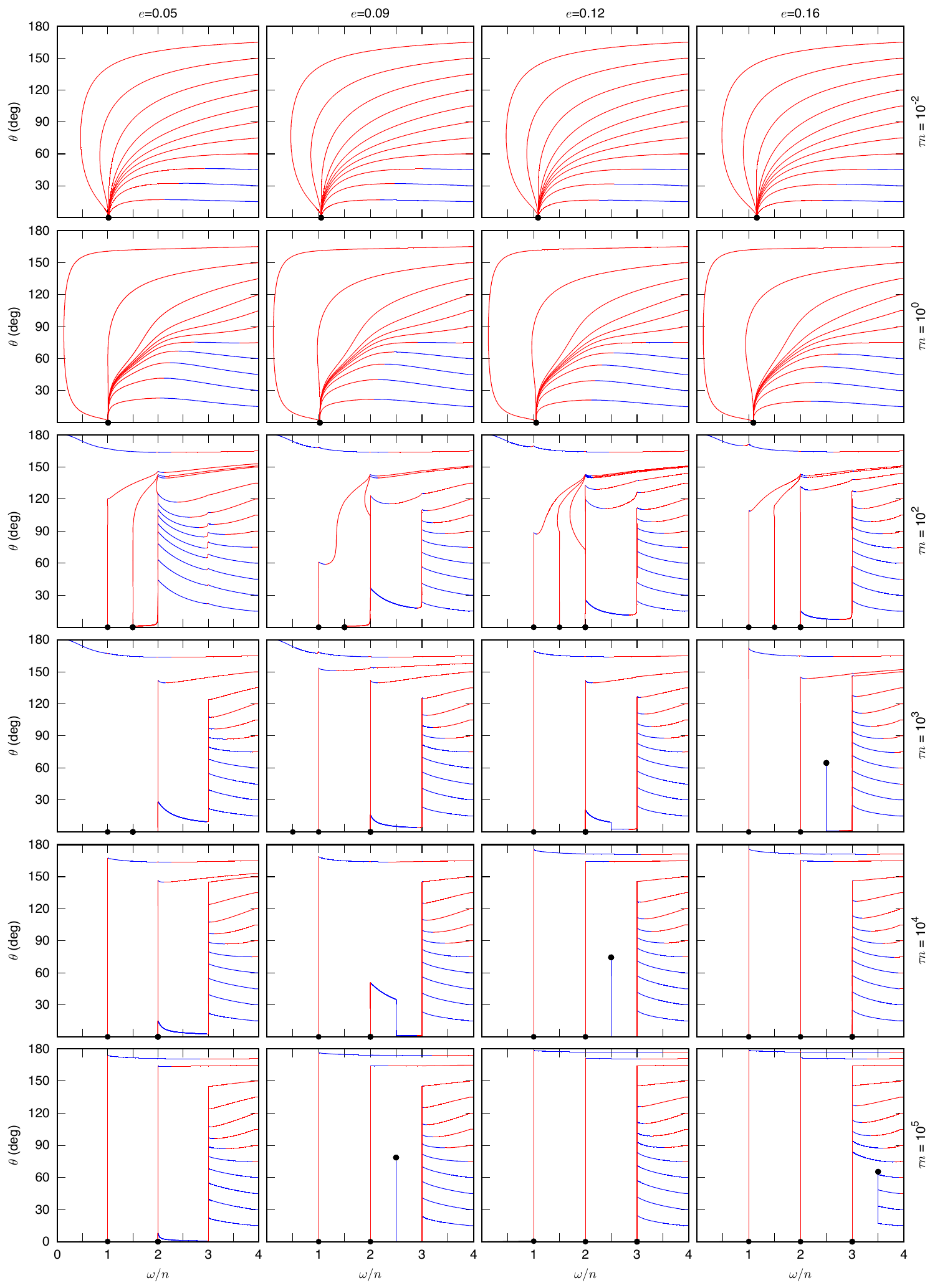}
\caption{Secular evolution of the spin in the plane ($\omega/n, \theta$) for different eccentricities and $\nt$-values. From left to right, the eccentricity increases from 0.05 to 0.16. From \bfx{top to bottom}, the product $\nt$ increases from $10^{-2}$ to $10^5$. This plot was obtained with initial $\om/n = 3.93$, and by integrating Eqs. \eqref{220407a} and \eqref{220407c}, where the Hansen coefficients $X_k^{-3,m} (e)$ are truncated at $|k|\leq 20$.
Segments with $\dot \theta < 0$ are plotted in red, while segments with $\dot \theta > 0$ are plotted in blue.
Fixed points are marked with a black dot. \label{FigI3}}
\end{figure*}

In Fig.~\ref{FigI3}, we show some trajectories for the spin in the plane $(\omega/n, \theta)$ for different values of $\nt$ (between $10^{-2}$ and $10^5$) and constant eccentricity\footnote{Fig.~\ref{FigI3} in this paper is similar to Fig.~2 in \citet{Boue_etal_2016}, who adopted $\nt \le 10^2$ and different eccentricities ($e=0.0$, $0.3$ and $0.6$). These choices did not allow them to spot some interesting behaviours, such as the stable non-zero obliquity equilibrium states.} 
(we adopt $0.05 \le e \le 0.16$, because close-in Earth-like planets around M-dwarfs are usually observed with small eccentricities).
We start the integrations with $\om/n = 3.93$ and different obliquity values.
As tidal effects initially decrease the rotation rate (Eq.\,\eqref{220419h}), the evolutionary paths must be followed for decreasing values of $\om/n$, until they reach an equilibrium position.
Final metastable states (i.e. fixed points for a constant eccentricity) are marked with a black dot.
To better understand the obliquity behaviour, segments with $\dot \theta < 0$ are plotted in red (the obliquity is damped), while segments with $\dot \theta > 0$ are plotted in blue (the obliquity grows). 

For $\nt \lesssim 1$, the system is in the linear regime. 
For initial small obliquities and $\om/n > 2$, the obliquity increases (Eq.\,\eqref{220419i}).
However, as the rotation rate decreases, $\theta=0$ becomes stable (Eq.\,\eqref{220419c}). 
There are no other stable equilibrium points for the obliquity in this regime, and therefore all trajectories evolve to zero obliquity  in the end. 
The final rotation rate is given by the pseudo-rotation equilibrium (Eq.\,\eqref{090520a}).

For $\nt \gg 1$, multiple resonant equilibria exist at $\om/n = k/2$ ($k \in \mathbb{Z}$), even for almost zero eccentricity.
In general, the number of possibilities increases with the eccentricity and $\nt$ (Fig.~\ref{FigI1}), but also with the obliquity (Eq.\,\eqref{211110t1}).
The resonant entrapments are well visible by vertical lines, because although the rotation is stalled, the obliquity continues to evolve.
Most of these lines are in red, meaning that the obliquity decreases for those resonances.
In some of them, the obliquity is able to reach $\theta=0$, where the rotation rate stabilises (black dot).
In other cases (e.g. for $\om/n=3$ with $\nt=10^2$), as the obliquity decreases, the resonance becomes unstable, and the rotation decreases into a lower order resonance.

As $\nt$ increases, more resonant equilibria become available (Fig.~\ref{FigI1}).
For $\nt \gtrsim 10^3$, the obliquity is able to grow at some resonances (blue vertical lines).
This behaviour is observed for $\om/n = 5/2$ and $7/2$, because the torque on the obliquity can be always positive for those resonances (see Sect.~\ref{oblexct} and Fig.~\ref{FigI2}).
In some cases, the high obliquity ends by destabilising the resonant equilibrium\footnote{A short-term small obliquity excitation in the $7/2$~resonance was originally reported by \citet{Boue_etal_2016} while using $\nt=10^2$ and $e=0.3$. The obliquity climbs during about 15~Myr up to $\theta\approx15^\circ$, after which the $7/2$~resonance becomes unstable.}, 
and the rotation rate decreases to a lower order spin--orbit resonance. 
More interestingly, however, we confirm that it is possible to stabilise the spin in a high-obliquity state (Fig.~\ref{eqobecc}).
In Fig.~\ref{FigI3}, these stable high-obliquity equilibria are observed for ($\om/n \simeq 5/2, \theta \simeq 65^\circ-80^\circ$) when 
$e=0.16$ and $\nt = 10^3$, 
$e=0.12$ and $\nt = 10^4$, or
$e=0.09$ and $\nt = 10^5$, 
and for ($\om/n \simeq 7/2, \theta \simeq 65^\circ$) when $e=0.16$ and $\nt = 10^5$.

As we start the simulations with $\om/n \approx 4$, most trajectories are initially captured in the nearby $7/2$ and $3/1$ spin--orbit resonances.
Nevertheless, lower order resonances can still be attained when the $3/1$~resonance becomes unstable for low obliquity.
Lower order resonances can also be directly attained for initial obliquities higher than $150^\circ$, although in this case the initial obliquity has to be finely adjusted.
For $\nt = 10^2$, we observe that initial obliquities higher than $165^\circ$ evolve into $180^\circ$, while the rotation rate becomes negative.
These simulations are continued at $\theta=0$ with the rotation rate increasing into the synchronous state, because from a dynamical point of view, the pair $(-\omega, \theta = \pi)$ is equivalent to the pair $(\omega, \theta = 0)$.

In Fig.~\ref{FigI3}, we observe that there is a large variety of evolutionary scenarios for the spin of close-in planets that depend on the initial obliquity, eccentricity, and relaxation time.
\bfx{For instance, for $e=0.09$ and $\nt=10^4$, trajectories with initial obliquity lower than $150^\circ$ are at first captured in the $3/1$ spin--orbit resonance, where the obliquity decreases until it reaches a value close to zero. 
At this point, the $3/1$~resonance becomes unstable, and the rotation decreases.
Subsequently, capture in the $5/2$~resonance occurs, but here the obliquity increases.
When the obliquity reaches about $40^\circ$, the $5/2$~resonance becomes unstable, and the rotation rate finally evolves into the $2/1$ spin--orbit resonance, where it stabilises with zero obliquity.}

The different evolution scenarios depicted in Fig.~\ref{FigI3} are completely general, as they do not depend on the semi-major axis, the radius, or the masses.
These parameters \bfx{appear} in $\At/L$ and thus only modify the spin evolution timescale (Eq.\,\eqref{220414a}).
In theory, for a given $\nt$-value, it is therefore possible to predict the present spin-state of a close-in Earth-like planet solely from knowledge of its current eccentricity.

The spin evolution is also affected by the orbital evolution, because the eccentricity is not constant (Eq.\,\eqref{220406d}).
The eccentricity generally decreases over a longer timescale, and so Fig.~\ref{FigI3} can still be used to predict the long-term evolution of the system.
Each panel can be seen as a snapshot at a given eccentricity, evolving from the right-hand side to the left.
As a result, the stable black dots only correspond to metastable states, because they disappear for decreasing eccentricity.
For small eccentricities only low-order spin--orbit resonances persist, and for zero eccentricity only the synchronous rotation is possible for all $\nt$-values \citep[see Fig.~5 in][]{Correia_etal_2014}.

\section{Application to Ross\,128\,b}
\label{apliRoss}

To get the complete spin and orbital evolution of a system \bfx{over time}, we need to attribute specific values to all orbital and physical parameters.
One good candidate for habitability studies is Ross\,128\,b, an exoplanet with a minimal mass of $\mp = 1.35\,M_\oplus$ \bfx{in a $9.86$-day orbit around a $\sim 5$~Gyr-old M-dwarf star with $\ms = 0.168\,M_\odot$} \citep{Bonfils_2018}.
Although the planet is much closer to its host star ($a\approx0.05$~au) than the Earth is to the Sun, because of the much smaller luminosity of M-dwarfs, it only receives about $1.4$ times as much flux as the Earth from the Sun.
Ross\,128\,b is therefore likely to be situated at the inner edge of the habitable zone \citep{Bonfils_2018, Souto_2018}. 
In this section, we adopt the parameters listed in Table~\ref{TabX2}, and follow the tidal evolution of this planet by integrating the full set of secular equations of motion \eqref{220406c}$-$\eqref{220407b} for 10~Gyr.

\begin{table}
\caption{Parameters of the Ross\,128 system \citep{Bonfils_2018}.  \label{TabX2} } 
\begin{center}
{ \begin{tabular}{cc} \hline \hline
Parameter & Value  \\ \hline
$\ms$ & $0.168\,M_\odot$ \\
$\mp$ & $1.35\,M_\oplus$ \\
$a$ & $0.0496$~au \\ 
age & $\gtrsim 5$~Gyr \\ \hline
$\Rp$ & $1.12\,R_\oplus$ \\
$\xi$ & $0.330$ \\
$\ke$ & $0.295$ \\
$\kf$ & $0.933$ \\ \hline
\end{tabular}}
\end{center}
$\Rp$ is estimated from $\mp$ assuming an Earth-like density. $\xi$, $\ke$ and $\kf$ are equal to the Earth's values \citep{Yoder_1995cnt}.
\end{table}

The initial spin state of Earth-like planets is unknown.
A small number of large impacts at the end of their formation can modify the rotation period and the axis orientation \citep{Dones_Tremaine_1993, Kokubo_Ida_2007}.
However, as planets contract from a larger protoplanet, in general it is expected that they rotate fast.
For simplicity, here we assume in all simulations that the initial rotation period of Ross\,128\,b is 24~h, which gives an initial rotation rate of $\om / n = 9.86$.
This value is not critical, because the rotation rate rapidly decreases and evolves into an equilibrium value (Sect.~\ref{eqrot}).
We also assume the same initial obliquity $\theta=10^\circ$  in all simulations, because obliquities smaller than $150^\circ$ are expected to behave in a similar way (Fig.~\ref{FigI3}).
Moreover, by adopting a small initial obliquity value, it is possible to make the obliquity \bfx{growth} stand out.

The semi-major axis does not change much over the course of the simulations, even for $\nt=10^2$.
Indeed, owing to the conservation of the orbital angular momentum (Eq.\,\eqref{210804a}), the final semi-major axis is given by $a_f \approx a (1-e^2)$, which results in a variation of only 4\% for $e=0.2$.
\bfx{In addition, the semi-major axis only impacts the evolution timescale through the parameters $\At$ and $\nG$ (Eq.\,\eqref{220414a}).}
Therefore, the choice of the initial semi-major axis is not critical, and so for simplicity we adopt the present value as the initial one.

The present eccentricity of Ross\,128\,b is not very well constrained from the observations, $e = 0.12 \pm 0.09$ \citep{Bonfils_2018}, that is, it is compatible with zero, but it can also be higher than 0.2.
However, this parameter is critical for the spin evolution of the planet (Fig.~\ref{FigI3}).
Therefore, we run our simulations adopting different values for the initial eccentricity from 0.01 to 0.20 in order to explore all the different behaviours.

Finally, \bfx{although} Ross\,128\,b is slightly more massive than the Earth, we assume a similar rheology for tides.
For the density, structure constant, and Love numbers, we \bfx{adopt} those from the Earth (Table~\ref{TabX2}).
As discussed in Sect.~\ref{tidalmodels}, the relaxation time is poorly constrained, but a typical adopted value is $\te \sim 400$~yr, which gives $\nt \sim 10^5$.
However, this $\nt$-value leads to almost no orbital evolution over gigayear timescales, and therefore in order to get a more complete view of the possible spin evolutions, we run our simulations using different values, from $\nt = 10^2$ to $10^5$.

The results of the simulations for the tidal evolution of Ross\,128\,b are plotted in Figs.~\ref{fig_nt2} to~\ref{fig_nt5}, one for each value of $\nt$.
We show the eccentricity (top), the $\om/n$ ratio (middle), and the obliquity (bottom) as a function of time for different initial eccentricities.
\bfx{For reasons of clarity}, we only show six trajectories per plot, which illustrate all the important observed evolutions.

In Fig.~\ref{fig_nt2}, we adopt $\nt=10^2$, which corresponds to the highest tidal dissipation, because $\dot \om \propto \tau^{-1}$ (Eq.\,\eqref{220418a}).
As a result, the rotation rate reaches the spin--orbit resonance regime in \bfx{less than} 1~Myr, and the orbit circularises in about 5~Gyr.
We zoom into the early evolution ($< 1$~Myr), for a clearer view of the initial spin evolution (left hand side of Fig.~\ref{fig_nt2}).
We observe that for initial $e=0.2$, the rotation rate is captured in the $5/2$~resonance, for initial $e=0.1$, it is captured in the $3/2$~resonance, while for the remaining eccentricities between these two, all trajectories were captured in the $2/1$~resonance.
As expected, higher initial eccentricities foster capture in higher order spin--orbit resonances (Fig.~\ref{FigI1}).
Before resonance entrapment, the obliquity increases for all simulations (Eq.\,\eqref{220419i}).
However, as soon as capture in resonance occurs, the obliquity drops quickly to zero, except for the $5/2$~resonance, where it continues to increase.
After the obliquity surpasses a certain value ($\sim 35^\circ$), 
the $5/2$~resonance becomes unstable, and the rotation decreases to the $2/1$~resonance, where the obliquity is brought to zero.
We limit the final evolution to 0.5~Gyr (right hand side of Fig.~\ref{fig_nt2}), because as the eccentricity decreases, all spin--orbit resonances become unstable one by one, until \bfx{the rotation finally synchronises}.
Beyond that date, the eccentricity decays exponentially to zero, while the rotation remains synchronous and the obliquity at zero degrees.

In Fig.~\ref{fig_nt3}, we adopt $\nt=10^3$.
We observe that, although the tidal dissipation is about ten times weaker than for the case with $\nt=10^2$ (Eq.\,\eqref{220418a}), the rotation rate evolution does not differ much.
It is initially trapped in some spin--orbit resonance, but as the eccentricity decreases, they all become unstable one by one.
All trajectories finally end in the synchronous resonance before 5~Gyr (the present estimated age of the star).
After that point, the eccentricity decreases very slowly, remaining above zero for at least 50~Gyr.
One difference with respect to the case with $\nt=10^2$ is that the resonance entrapment persists for a much longer time, \bfx{because the eccentricity is also damped much more slowly.}
Another difference is that for the trajectories captured in the $5/2$~resonance, the obliquity is able to grow to a steady threshold around $65^\circ$ (Fig.~\ref{eqobecc}).
As the eccentricity decreases, the equilibrium obliquity slightly increases, until the $5/2$~resonance is destabilised.
At this stage, the rotation rate decreases until it is captured in the lower order $2/1$~resonance and the obliquity damps to zero.
We observe that the obliquity growth in the $5/2$~resonance is very consistent.
For initial eccentricities higher than 0.16, the rotation is initially captured in the $3/1$~resonance and the obliquity becomes very close to zero.
However, as the $3/1$~resonance becomes unstable, the rotation rate is trapped in the $5/2$~resonance, where the obliquity rises again and settles in the $65^\circ$ threshold.
The high obliquity state is maintained as long as the eccentricity is above its critical value $e\approx 0.15$ for $\nt=10^3$ (Fig.~\ref{eqobecc}).

\begin{figure}
\centering
\includegraphics[width=\columnwidth]{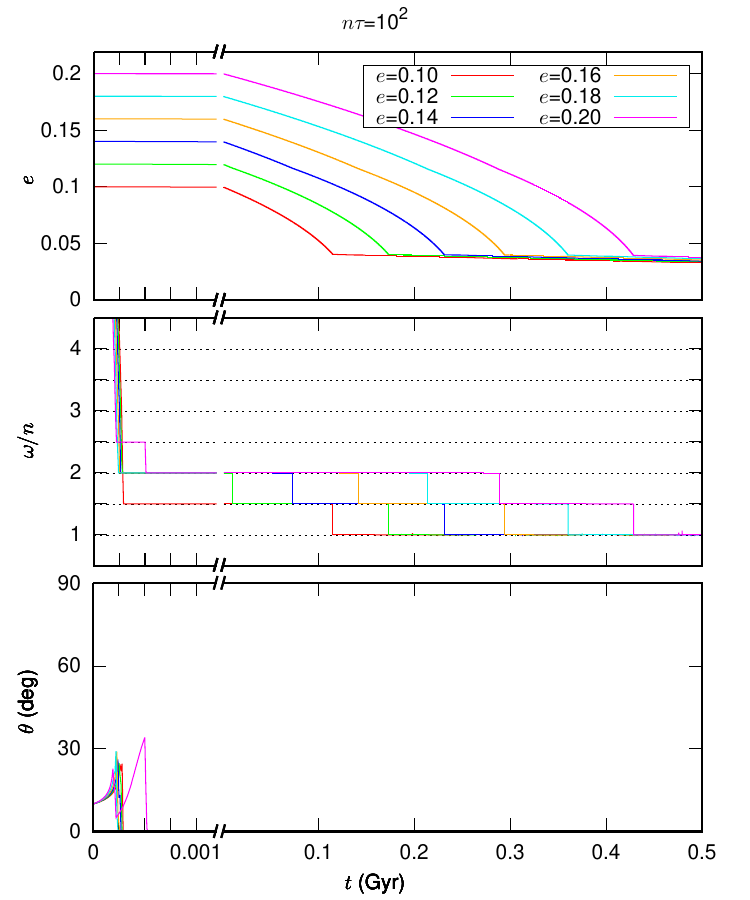}
\caption{Secular evolution \bfx{over time} of Ross\,128\,b with $\nt=10^2$ and some initial eccentricity values between $0.10$ and $0.20$. We show the eccentricity (top), the $\om/n$ ratio (middle) and the obliquity (bottom).
In all simulations, the initial rotation period is 24~h ($\om/n = 9.86$), and the initial obliquity is $\theta=10^\circ$. \bfx{We note the break in the timescale.}  \llabel{fig_nt2}}
\end{figure}

\begin{figure}
\centering
\includegraphics[width=\columnwidth]{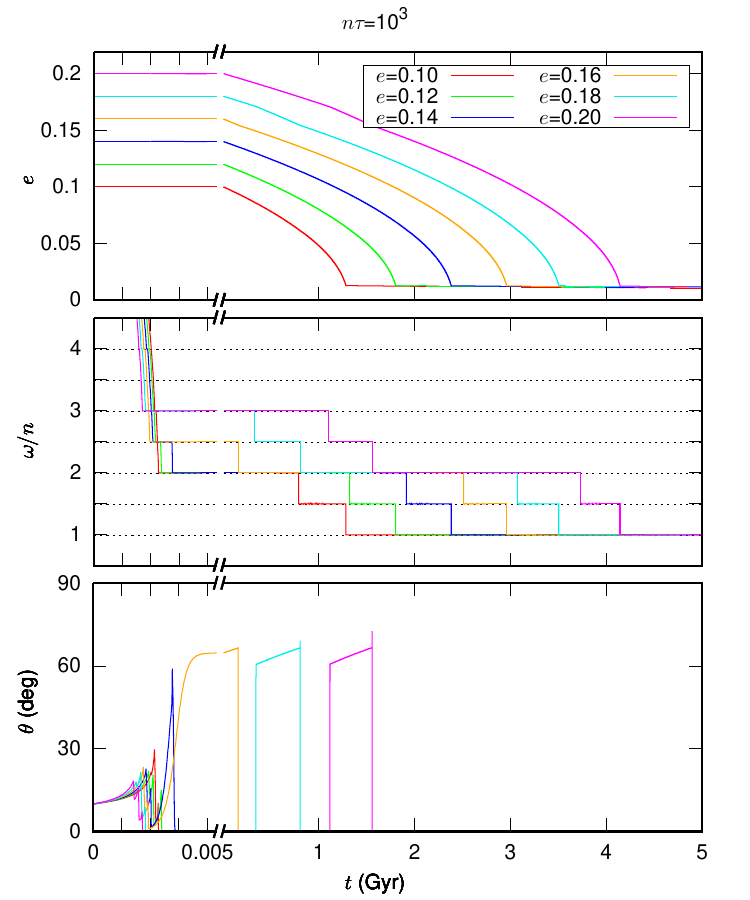}
\caption{Secular evolution \bfx{over time} of Ross\,128\,b with $\nt=10^3$ and some initial eccentricity values between $0.10$ and $0.20$. We show the eccentricity (top), the $\om/n$ ratio (middle) and the obliquity (bottom).
In all simulations, the initial rotation period is 24~h ($\om/n = 9.86$), and the initial obliquity is $\theta=10^\circ$. \bfx{We note the break in the timescale.}  \llabel{fig_nt3}}
\end{figure}

\begin{figure}
\centering
\includegraphics[width=\columnwidth]{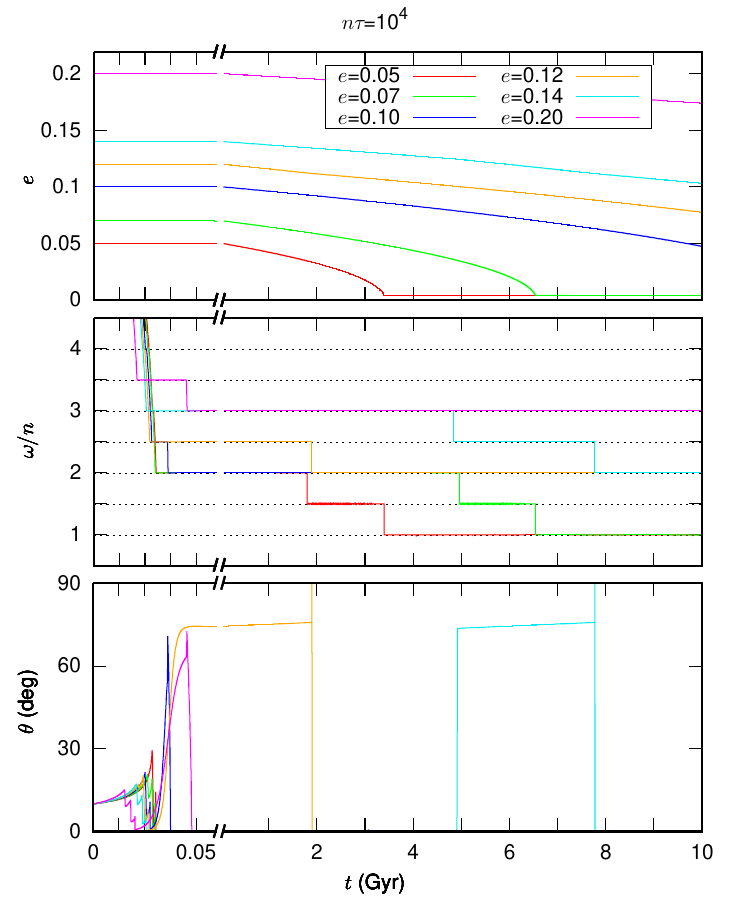}
\caption{Secular evolution \bfx{over time} of Ross\,128\,b with $\nt=10^4$ and some initial eccentricity values between $0.05$ and $0.20$. We show the eccentricity (top), the $\om/n$ ratio (middle) and the obliquity (bottom).
In all simulations, the initial rotation period is 24~h ($\om/n = 9.86$), and the initial obliquity is $\theta=10^\circ$. \bfx{We note the break in the timescale.}  \llabel{fig_nt4}}
\end{figure}

\begin{figure}
\centering
\includegraphics[width=\columnwidth]{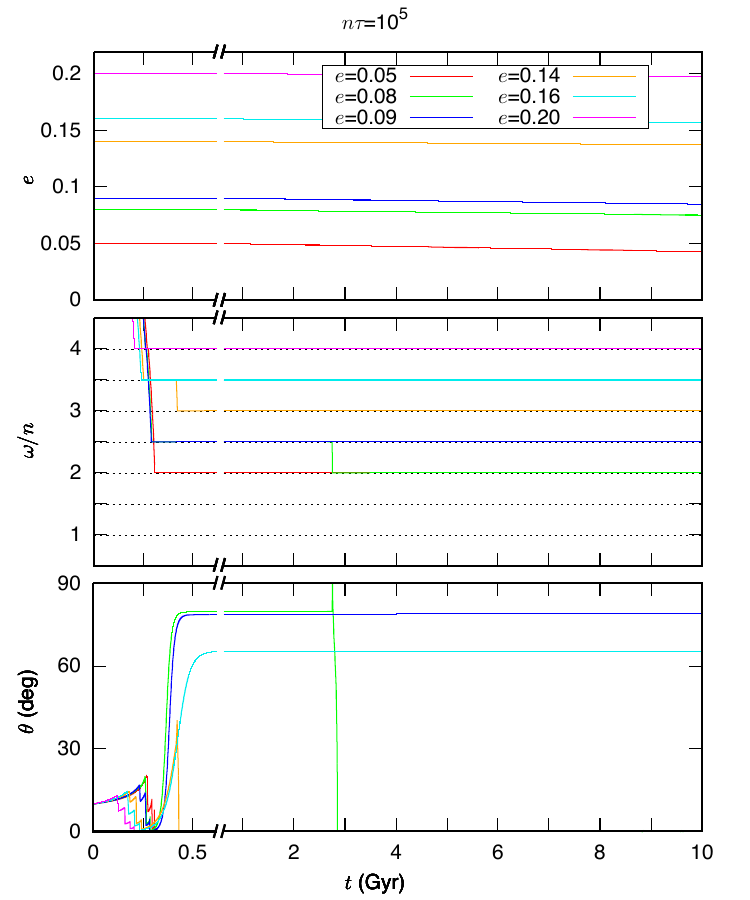}
\caption{Secular evolution \bfx{over time} of Ross\,128\,b with $\nt=10^5$ and some initial eccentricity values between $0.05$ and $0.20$. We show the eccentricity (top), the $\om/n$ ratio (middle) and the obliquity (bottom).
In all simulations, the initial rotation period is 24~h ($\om/n = 9.86$), and the initial obliquity is $\theta=10^\circ$. \bfx{We note the break in the timescale.}  \label{fig_nt5}}
\end{figure}

In Fig.~\ref{fig_nt4}, we adopt $\nt=10^4$.
The evolution of the spin is similar to the previous case with $\nt=10^3$, but the evolution timescale is about ten times longer.
The eccentricity therefore varies very slowly and does not change much over the 10~Gyr of the simulation.
As a consequence, capture in each spin--orbit resonance can last for several gigayears and only trajectories with initial $e < 0.1$ evolve into the synchronous rotation before 10~Gyr.
For trajectories captured in the $5/2$~resonance, the obliquity grows to nearly $75^\circ$ and can remain there up to 3~Gyr.
These high-obliquity states are very robust, as they correspond to \bfx{fixed points} and can only be destabilised when the eccentricity descends below the critical value of $e\approx 0.11$ (Fig.~\ref{eqobecc}).
Large $\nt$-values permit capture in high-order spin--orbit resonances (Fig.~\ref{FigI1}), which includes the $7/2$~resonance when $\nt=10^4$.
In this resonance, the obliquity is also allowed to grow to high values.
In our simulations, this scenario is observed for initial $e=0.20$, but the obliquity cannot stabilise, because the critical eccentricity for this resonance is $e\approx 0.21$ for $\nt=10^4$. 

In Fig.~\ref{fig_nt5}, we finally adopt $\nt=10^5$, which corresponds to the more realistic value of $\nt$ for Earth-like planets (see Sect.~\ref{tidalmodels}).
In this case, tidal dissipation is so weak that the initial eccentricity almost does not change over the 10~Gyr of the simulation.
As a result, the evolution of the spin can be depicted from the general evolution map shown in the last row of Fig.~\ref{FigI3}.
Indeed, most trajectories remain throughout the simulation in the spin--orbit resonance where they were initially captured, \bfx{except} for a few cases ($e=0.08$ and $e=0.14$).
For the trajectories that were captured in the $5/2$ and $7/2$~resonance, the obliquity is allowed to grow and stabilise at $79^\circ$ and $66^\circ$, respectively (Fig.~\ref{eqobecc}).
For $\nt=10^5$, we also observe one capture in the higher order $4/1$ spin--orbit resonance for $e=0.2$.
This trajectory is not covered by the possibilities shown in Fig.~\ref{FigI3}, but in the 4/1~resonance the obliquity also evolves to zero.

Evolution trajectories for $\nt > 10^5$ are not shown, because the eccentricity is already nearly constant for $\nt=10^5$.
We expect that the number of spin--orbit resonances increases for larger values of $\nt$ (beyond the $4/1$~resonance), which remain stable throughout the entire lifetime of the system.
High-obliquity states are also possible and persist for long timescales, as they require lower eccentricities to become unstable (Fig.~\ref{eqobecc}).

In this section, we present the results for Ross\,128\,b, but we also run simulations for other Earth-like planets in the habitable zone of M-dwarf stars, such as Proxima\,b \citep{Anglada_Escude_etal_2016} and Teegarden's\,c \citep{Zechmeister_etal_2019}.
We observed similar results, which confirms that the general analysis described in Sect.~\ref{obli} is robust and independent from a fine tuning of the system parameters.

\section{Discussion and Conclusions}
\label{conclusion}

Earth-like rocky planets in the habitable zone of M-dwarfs, such as Ross\,128\,b, are expected to present a viscoelastic rheology with relaxation times such that $\nt \gg 1$.
In this regime, a multitude of spin--orbit resonances arise, where the spin can be temporarily trapped.
Large $\nt$-values increase the number of possible resonances, but also their lifetime, because the eccentricity damping timescale is longer.
In addition, for the $5/2$ and $7/2$ resonances, the obliquity can grow to high values of between $60^\circ$ and $80^\circ$, which can be maintained for several gigayears. 
These results are extremely important regarding the habitability of these planets, because the rotation period and the obliquity determine the insolation distribution on the surface of the planet \citep[e.g.][]{Milankovitch_1941}.
The rotation rate controls the length of the day--night cycle through the synodic day period:
\be
P_\mathrm{syn} = \frac{2 \pi}{\om - n} = \frac{P_\mathrm{orb}}{\om/n -1} 
\ . \llabel{220428a}
\ee
For synchronous rotation ($\om/n = 1$), we have $P_\mathrm{syn}  = \infty$, and so one side of the planet always faces the star and becomes extremely hot, while the other side becomes very cold.
Therefore, synchronous rotation is usually evoked as one of the major obstacles for the habitability of planets around M-dwarfs.
However, in a $2/1$~resonance, we have $P_\mathrm{syn}=P_\mathrm{orb}$, which corresponds to $P_\mathrm{syn} = 9.86$~day in the case of Ross\,128\,b, while for the $7/2$~resonance, we get $P_\mathrm{syn}= 3.94$~day.
The obliquity impacts the insolation distribution for a given latitude.
It is responsible for the seasons, and therefore, even for synchronous planets, high obliquities can also mimic day--night cycles with the duration of one year \citep{Dobrovolskis_2009}, which corresponds to 9.86~days in the case of Ross\,128\,b.
The combination of asynchronous rotation with extreme obliquities may thus help to sustain temperate environments and more favourable conditions for life.

Our results were obtained with a very general formalism, and are therefore not restricted to the study of rocky planets.
They can be extended to other tidal regimes (different $\nt$-values) or to other rheologies by simply modifying the Love number (Eq.\,\eqref{210929a}).
In our analysis (Sect.~\ref{obli}), we already considered relaxation times such that $\nt \lesssim 1$, which are expected for gaseous planets \bfx{\citep[e.g.][]{Correia_etal_2014}}.
In this regime, the Maxwell model behaves as the linear model, and the results are well known from previous studies \citep[e.g.][]{Correia_Laskar_2010}: the rotation rate evolves to the pseudo-synchronous value (Eq.\,\eqref{090520a}), while the obliquity evolves to zero.
These results are less interesting from a habitability point of view, but can still be used to put constraints on global atmospheric circulation models for planets in this regime, such as \bfx{hot Jupiters} or \bfx{warm Neptunes}.

In the two-body problem, tides damp the eccentricity, which in turn destabilises the different equilibria for the spin.
However, low-mass planets in the habitable zone of M-dwarfs are usually found in multi-planet systems \citep[e.g.][]{Dressing_Charbonneau_2015}.
As a result, the eccentricity can be excited by mutual perturbations and keep a non-zero value \bfx{despite} the tidal dissipation \citep{Laskar_etal_2012, Barnes_2017}.
In multi-planet systems, the obliquity can also be locked in a Cassini state where it retains a non-zero value \citep{Colombo_1966}.
Strong tidal effects usually destabilise high-obliquity Cassini states \citep[e.g.][]{Levrard_etal_2007}, but in the peculiar $5/2$ and $7/2$~resonances the tidal torque can help to sustain those equilibria.

The tidal equilibria for the spin ($\om/n, \theta$) depends only on two key parameters: 
the eccentricity, $e$, and the relative relaxation time, $\tau$.
The eccentricity is a parameter that we can already access with the current observational techniques, in particular for planets detected with the radial-velocity method.
$\tau$ is therefore the only totally unknown parameter, and needs to be explored in order to put constraints on the possible spin state of a close-in planet.
Inversely, if we are able to measure the spin of a given planet, we can put constraints on $\tau$.
Although not possible at present, a new generation of instruments, such as the ELT-METIS integral field spectrograph \citep{Quanz_etal_2015, Brandl_etal_2021}, will combine high-contrast imaging with high-resolution spectroscopy, and allow us to determine the spin state of these planets.

\begin{acknowledgements}
\bfx{We are grateful to the referee Anthony Dobrovolskis for helpful comments.}
This work was supported by
CFisUC (UIDB/04564/2020 and UIDP/04564/2020),
GRAVITY (PTDC/FIS-AST/7002/2020),
PHOBOS (POCI-01-0145-FEDER-029932), and
ENGAGE~SKA (POCI-01-0145-FEDER-022217),
funded by COMPETE 2020 and FCT, Portugal.
\bfx{The numerical simulations were performed at the Laboratory for Advanced Computing at University of Coimbra (\href{https://www.uc.pt/lca}{https://www.uc.pt/lca}).}
\end{acknowledgements}

\bibliographystyle{aa}
\bibliography{references}

\begin{thebibliography}{74}
\expandafter\ifx\csname natexlab\endcsname\relax\def\natexlab#1{#1}\fi

\bibitem[{{Adams} \& {Bloch}(2015)}]{Adams_Bloch_2015}
{Adams}, F.~C. \& {Bloch}, A.~M. 2015, \mnras, 446, 3676

\bibitem[{{Anglada-Escud{\'e}} {et~al.}(2016){Anglada-Escud{\'e}}, {Amado},
  {Barnes}, {Berdi{\~n}as}, {Butler}, {Coleman}, {de La Cueva}, {Dreizler},
  {Endl}, {Giesers}, {Jeffers}, {Jenkins}, {Jones}, {Kiraga}, {K{\"u}rster},
  {L{\'o}pez-Gonz{\'a}lez}, {Marvin}, {Morales}, {Morin}, {Nelson}, {Ortiz},
  {Ofir}, {Paardekooper}, {Reiners}, {Rodr{\'\i}guez},
  {Rodr{\'\i}guez-L{\'o}pez}, {Sarmiento}, {Strachan}, {Tsapras}, {Tuomi}, \&
  {Zechmeister}}]{Anglada_Escude_etal_2016}
{Anglada-Escud{\'e}}, G., {Amado}, P.~J., {Barnes}, J., {et~al.} 2016, \nat,
  536, 437

\bibitem[{Anglada-Escudé {et~al.}(2013)Anglada-Escudé, Tuomi, Gerlach,
  Barnes, Heller, Jenkins, Wende, Vogt, Paul~Butler, Reiners, \&
  et~al.}]{Anglada_Escude_2013}
Anglada-Escudé, G., Tuomi, M., Gerlach, E., {et~al.} 2013, Astronomy \&
  Astrophysics, 556, A126

\bibitem[{Armstrong {et~al.}(2014)Armstrong, Barnes, Domagal-Goldman, Breiner,
  Quinn, \& Meadows}]{Armstrong_2014}
Armstrong, J., Barnes, R., Domagal-Goldman, S., {et~al.} 2014, Astrobiology,
  14, 277–291

\bibitem[{Barnes(2017)}]{Barnes_2017}
Barnes, R. 2017, Celestial Mechanics and Dynamical Astronomy, 129, 509–536

\bibitem[{Bochanski {et~al.}(2010)Bochanski, Hawley, Covey, West, Reid,
  Golimowski, \& Ivezić}]{Bochanski_2010}
Bochanski, J.~J., Hawley, S.~L., Covey, K.~R., {et~al.} 2010, The Astronomical
  Journal, 139, 2679–2699

\bibitem[{Bolmont {et~al.}(2016)Bolmont, Libert, Leconte, \&
  Selsis}]{Bolmont_2016}
Bolmont, E., Libert, A.-S., Leconte, J., \& Selsis, F. 2016, Astronomy \&
  Astrophysics, 591, A106

\bibitem[{{Bonfils} {et~al.}(2018){Bonfils}, {Astudillo-Defru}, {D{\'\i}az},
  {Almenara}, {Forveille}, {Bouchy}, {Delfosse}, {Lovis}, {Mayor}, {Murgas},
  {Pepe}, {Santos}, {S{\'e}gransan}, {Udry}, \& {W{\"u}nsche}}]{Bonfils_2018}
{Bonfils}, X., {Astudillo-Defru}, N., {D{\'\i}az}, R., {et~al.} 2018, \aap,
  613, A25

\bibitem[{{Bou{\'e}} {et~al.}(2016){Bou{\'e}}, {Correia}, \&
  {Laskar}}]{Boue_etal_2016}
{Bou{\'e}}, G., {Correia}, A.~C.~M., \& {Laskar}, J. 2016, Celestial Mechanics
  and Dynamical Astronomy, 126, 31

\bibitem[{{Brandl} {et~al.}(2021){Brandl}, {Bettonvil}, {van Boekel},
  {Glauser}, {Quanz}, {Absil}, {Amorim}, {Feldt}, {Glasse}, {G{\"u}del}, {Ho},
  {Labadie}, {Meyer}, {Pantin}, {van Winckel}, \& {METIS
  Consortium}}]{Brandl_etal_2021}
{Brandl}, B., {Bettonvil}, F., {van Boekel}, R., {et~al.} 2021, The Messenger,
  182, 22

\bibitem[{{Burn} {et~al.}(2021){Burn}, {Schlecker}, {Mordasini}, {Emsenhuber},
  {Alibert}, {Henning}, {Klahr}, \& {Benz}}]{Burn_etal_2021}
{Burn}, R., {Schlecker}, M., {Mordasini}, C., {et~al.} 2021, \aap, 656, A72

\bibitem[{{Colombo}(1966)}]{Colombo_1966}
{Colombo}, G. 1966, \aj, 71, 891

\bibitem[{Colose {et~al.}(2019)Colose, Genio, \& Way}]{Colose_2019}
Colose, C.~M., Genio, A. D.~D., \& Way, M.~J. 2019, The Astrophysical Journal,
  884, 138

\bibitem[{{Correia}(2006)}]{Correia_2006}
{Correia}, A.~C.~M. 2006, \epsl, 252, 398

\bibitem[{{Correia}(2009)}]{Correia_2009}
{Correia}, A.~C.~M. 2009, \apjl, 704, L1

\bibitem[{{Correia}(2015)}]{Correia_2015}
{Correia}, A.~C.~M. 2015, \aap, 582, A69

\bibitem[{{Correia} {et~al.}(2014){Correia}, {Bou{\'e}}, {Laskar}, \&
  {Rodr{\'{\i}}guez}}]{Correia_etal_2014}
{Correia}, A.~C.~M., {Bou{\'e}}, G., {Laskar}, J., \& {Rodr{\'{\i}}guez}, A.
  2014, \aap, 571, A50

\bibitem[{{Correia} \& {Delisle}(2019)}]{Correia_Delisle_2019}
{Correia}, A. C.~M. \& {Delisle}, J.-B. 2019, \aap, 630, A102

\bibitem[{{Correia} \& {Laskar}(2001)}]{Correia_Laskar_2001}
{Correia}, A.~C.~M. \& {Laskar}, J. 2001, \nat, 411, 767

\bibitem[{{Correia} \& {Laskar}(2010)}]{Correia_Laskar_2010}
{Correia}, A.~C.~M. \& {Laskar}, J. 2010, \icarus, 205, 338

\bibitem[{{Correia} \& {Rodr{\'{\i}}guez}(2013)}]{Correia_Rodriguez_2013}
{Correia}, A.~C.~M. \& {Rodr{\'{\i}}guez}, A. 2013, \apj, 767, 128

\bibitem[{{Correia} \& {Valente}(2022)}]{Correia_Valente_2022}
{Correia}, A. C.~M. \& {Valente}, E. F.~S. 2022, Celestial Mechanics and
  Dynamical Astronomy, 134, 24

\bibitem[{{Darwin}(1879)}]{Darwin_1879a}
{Darwin}, G.~H. 1879, Philos. Trans. R. Soc. London, 170, 1

\bibitem[{{Darwin}(1908)}]{Darwin_1908}
{Darwin}, G.~H. 1908, {Scientific Papers} (Cambridge University Press)

\bibitem[{Dittmann {et~al.}(2017)Dittmann, Irwin, Charbonneau, Bonfils,
  Astudillo-Defru, Haywood, Berta-Thompson, Newton, Rodriguez, Winters, \&
  et~al.}]{Dittmann_2017}
Dittmann, J.~A., Irwin, J.~M., Charbonneau, D., {et~al.} 2017, Nature, 544,
  333–336

\bibitem[{{Dobrovolskis}(2007)}]{Dobrovolskis_2007}
{Dobrovolskis}, A.~R. 2007, \icarus, 192, 1

\bibitem[{{Dobrovolskis}(2009)}]{Dobrovolskis_2009}
{Dobrovolskis}, A.~R. 2009, \icarus, 204, 1

\bibitem[{{Dobrovolskis}(2021)}]{Dobrovolskis_2021}
{Dobrovolskis}, A.~R. 2021, \icarus, 363, 114297

\bibitem[{{Dones} \& {Tremaine}(1993)}]{Dones_Tremaine_1993}
{Dones}, L. \& {Tremaine}, S. 1993, \icarus, 103, 67

\bibitem[{{Dressing} \& {Charbonneau}(2015)}]{Dressing_Charbonneau_2015}
{Dressing}, C.~D. \& {Charbonneau}, D. 2015, \apj, 807, 45

\bibitem[{{Efroimsky}(2012)}]{Efroimsky_2012b}
{Efroimsky}, M. 2012, Celestial Mechanics and Dynamical Astronomy, 112, 283

\bibitem[{{Ferraz-Mello}(2013)}]{Ferraz-Mello_2013}
{Ferraz-Mello}, S. 2013, Celestial Mechanics and Dynamical Astronomy, 116, 109

\bibitem[{{Ferraz-Mello}(2015)}]{Ferraz-Mello_2015b}
{Ferraz-Mello}, S. 2015, Celestial Mechanics and Dynamical Astronomy, 122, 359

\bibitem[{{France} {et~al.}(2020){France}, {Duvvuri}, {Egan}, {Koskinen},
  {Wilson}, {Youngblood}, {Froning}, {Brown}, {Alvarado-G{\'o}mez},
  {Berta-Thompson}, {Drake}, {Garraffo}, {Kaltenegger}, {Kowalski}, {Linsky},
  {Loyd}, {Mauas}, {Miguel}, {Pineda}, {Rugheimer}, {Schneider}, {Tian}, \&
  {Vieytes}}]{France_etal_2020}
{France}, K., {Duvvuri}, G., {Egan}, H., {et~al.} 2020, \aj, 160, 237

\bibitem[{Gillon {et~al.}(2017)Gillon, Triaud, Demory, Jehin, Agol, Deck,
  Lederer, de~Wit, Burdanov, Ingalls, \& et~al.}]{Gillon_2017}
Gillon, M., Triaud, A. H. M.~J., Demory, B.-O., {et~al.} 2017, Nature, 542,
  456–460

\bibitem[{{Goldreich} \& {Peale}(1966)}]{Goldreich_Peale_1966}
{Goldreich}, P. \& {Peale}, S. 1966, \aj, 71, 425

\bibitem[{{Hut}(1980)}]{Hut_1980}
{Hut}, P. 1980, \aap, 92, 167

\bibitem[{{Karato} \& {Wu}(1993)}]{Karato_Wu_1993}
{Karato}, S.-I. \& {Wu}, P. 1993, Science, 260, 771

\bibitem[{Kasting {et~al.}(1993)Kasting, Whitmire, \&
  Reynolds}]{Kasting_etal_1993}
Kasting, J.~F., Whitmire, D.~P., \& Reynolds, R.~T. 1993, Icarus, 101, 108

\bibitem[{{Kirby} \& {Kronenberg}(1987)}]{Kirby_Kronenberg_1987}
{Kirby}, S.~H. \& {Kronenberg}, A.~K. 1987, Reviews of Geophysics, 25, 1219

\bibitem[{{Kite} \& {Barnett}(2020)}]{Kite_Barnett_2020}
{Kite}, E.~S. \& {Barnett}, M.~N. 2020, Proceedings of the National Academy of
  Science, 117, 18264

\bibitem[{{Kokubo} \& {Ida}(2007)}]{Kokubo_Ida_2007}
{Kokubo}, E. \& {Ida}, S. 2007, \apj, 671, 2082

\bibitem[{kumar Kopparapu {et~al.}(2019)kumar Kopparapu, Wolf, \&
  Meadows}]{Kopparapu_2019}
kumar Kopparapu, R., Wolf, E.~T., \& Meadows, V.~S. 2019, Characterizing
  Exoplanet Habitability

\bibitem[{{Lambeck}(1980)}]{Lambeck_1980}
{Lambeck}, K. 1980, {The Earth's Variable Rotation: Geophysical Causes and
  Consequences} (Cambridge University Press)

\bibitem[{Lammer {et~al.}(2007)Lammer, Lichtenegger, Kulikov, Griessmeier,
  Terada, Erkaev, Biernat, Khodachenko, Ribas, Penz, \&
  Selsis}]{Lammer_etal_2007}
Lammer, H., Lichtenegger, H., Kulikov, Y., {et~al.} 2007, Astrobiology, 7, 185

\bibitem[{{Laskar} {et~al.}(2012){Laskar}, {Bou{\'e}}, \&
  {Correia}}]{Laskar_etal_2012}
{Laskar}, J., {Bou{\'e}}, G., \& {Correia}, A.~C.~M. 2012, \aap, 538, A105

\bibitem[{{Laskar} \& {Robutel}(1993)}]{Laskar_Robutel_1993}
{Laskar}, J. \& {Robutel}, P. 1993, \nat, 361, 608

\bibitem[{{Levrard} {et~al.}(2007){Levrard}, {Correia}, {Chabrier}, {Baraffe},
  {Selsis}, \& {Laskar}}]{Levrard_etal_2007}
{Levrard}, B., {Correia}, A.~C.~M., {Chabrier}, G., {et~al.} 2007, \aap, 462,
  L5

\bibitem[{{Makarov} {et~al.}(2012){Makarov}, {Berghea}, \&
  {Efroimsky}}]{Makarov_etal_2012}
{Makarov}, V.~V., {Berghea}, C., \& {Efroimsky}, M. 2012, \apj, 761, 83

\bibitem[{{Mignard}(1979)}]{Mignard_1979}
{Mignard}, F. 1979, Moon and Planets, 20, 301

\bibitem[{{Milankovitch}(1941)}]{Milankovitch_1941}
{Milankovitch}, M. 1941, {Kanon der Erdbestrahlung und seine Andwendung auf das
  Eiszeitenproblem} (Royal Serbian Academy)

\bibitem[{{Millholland} \& {Laughlin}(2019)}]{Millholland_Laughlin_2019}
{Millholland}, S. \& {Laughlin}, G. 2019, Nature Astronomy, 3, 424

\bibitem[{{Munk} \& {MacDonald}(1960)}]{Munk_MacDonald_1960}
{Munk}, W.~H. \& {MacDonald}, G.~J.~F. 1960, {The Rotation of the Earth; A
  Geophysical Discussion} (Cambridge University Press)

\bibitem[{{Murray} \& {Dermott}(1999)}]{Murray_Dermott_1999}
{Murray}, C.~D. \& {Dermott}, S.~F. 1999, {Solar System Dynamics} (Cambridge
  University Press)

\bibitem[{{Quanz} {et~al.}(2015){Quanz}, {Crossfield}, {Meyer}, {Schmalzl}, \&
  {Held}}]{Quanz_etal_2015}
{Quanz}, S.~P., {Crossfield}, I., {Meyer}, M.~R., {Schmalzl}, E., \& {Held}, J.
  2015, International Journal of Astrobiology, 14, 279

\bibitem[{Reiners {et~al.}(2018)Reiners, Ribas, Zechmeister, Caballero,
  Trifonov, Dreizler, Morales, Tal-Or, Lafarga, Quirrenbach, \&
  et~al.}]{Reiners_2018}
Reiners, A., Ribas, I., Zechmeister, M., {et~al.} 2018, Astronomy \&
  Astrophysics, 609, L5

\bibitem[{{Remus} {et~al.}(2012){Remus}, {Mathis}, \&
  {Zahn}}]{Remus_etal_2012b}
{Remus}, F., {Mathis}, S., \& {Zahn}, J.-P. 2012, \aap, 544, A132

\bibitem[{{Renaud} \& {Henning}(2018)}]{Renaud_Henning_2018}
{Renaud}, J.~P. \& {Henning}, W.~G. 2018, \apj, 857, 98

\bibitem[{{Schlecker} {et~al.}(2021){Schlecker}, {Pham}, {Burn}, {Alibert},
  {Mordasini}, {Emsenhuber}, {Klahr}, {Henning}, \&
  {Mishra}}]{Schlecker_etal_2021}
{Schlecker}, M., {Pham}, D., {Burn}, R., {et~al.} 2021, \aap, 656, A73

\bibitem[{{Schlichting}(2018)}]{Schlichting_2018}
{Schlichting}, H.~E. 2018, in Handbook of Exoplanets, ed. H.~J. {Deeg} \& J.~A.
  {Belmonte}, 141

\bibitem[{Shields {et~al.}(2016)Shields, Ballard, \& Johnson}]{Shields_2016}
Shields, A.~L., Ballard, S., \& Johnson, J.~A. 2016, Physics Reports, 663,
  1–38

\bibitem[{Souto {et~al.}(2018)Souto, Unterborn, Smith, Cunha, Teske, Covey,
  Rojas-Ayala, Garc{\'{\i}}a-Hern{\'{a}}ndez, Stassun, Zamora, Masseron,
  Johnson, Majewski, Jönsson, Gilhool, Blake, \& Santana}]{Souto_2018}
Souto, D., Unterborn, C.~T., Smith, V.~V., {et~al.} 2018, The Astrophysical
  Journal, 860, L15

\bibitem[{{Su} \& {Lai}(2022{\natexlab{a}})}]{Su_Lai_2022b}
{Su}, Y. \& {Lai}, D. 2022{\natexlab{a}}, \mnras, 513, 3302

\bibitem[{{Su} \& {Lai}(2022{\natexlab{b}})}]{Su_Lai_2022a}
{Su}, Y. \& {Lai}, D. 2022{\natexlab{b}}, \mnras, 509, 3301

\bibitem[{Tuomi {et~al.}(2014)Tuomi, Jones, Barnes, Anglada-Escudé, \&
  Jenkins}]{Tuomi_2014}
Tuomi, M., Jones, H. R.~A., Barnes, J.~R., Anglada-Escudé, G., \& Jenkins,
  J.~S. 2014, Monthly Notices of the Royal Astronomical Society, 441,
  1545–1569

\bibitem[{Tuomi {et~al.}(2019)Tuomi, Jones, Butler, Arriagada, Vogt, Burt,
  Laughlin, Holden, Shectman, Crane, Thompson, Keiser, Jenkins, Berdiñas,
  Diaz, Kiraga, \& Barnes}]{Tuomi_etal_2019}
Tuomi, M., Jones, H. R.~A., Butler, R.~P., {et~al.} 2019, Frequency of planets
  orbiting M dwarfs in the Solar neighbourhood

\bibitem[{Turbet {et~al.}(2016)Turbet, Leconte, Selsis, Bolmont, Forget, Ribas,
  Raymond, \& Anglada-Escudé}]{Turbet_2016}
Turbet, M., Leconte, J., Selsis, F., {et~al.} 2016, Astronomy \& Astrophysics,
  596, A112

\bibitem[{{Turcotte} \& {Schubert}(2002)}]{Turcotte_Schubert_2002}
{Turcotte}, D.~L. \& {Schubert}, G. 2002, {Geodynamics}

\bibitem[{{Ward}(1974)}]{Ward_1974}
{Ward}, W.~R. 1974, \jgr, 79, 3375

\bibitem[{{Winn}(2018)}]{Winn_2018}
{Winn}, J.~N. 2018, in Handbook of Exoplanets, ed. H.~J. {Deeg} \& J.~A.
  {Belmonte}, 195

\bibitem[{Wordsworth(2015)}]{Wordsworth_2015}
Wordsworth, R. 2015, The Astrophysical Journal, 806, 180

\bibitem[{Yang {et~al.}(2014)Yang, Boué, Fabrycky, \& Abbot}]{Yang_2014}
Yang, J., Boué, G., Fabrycky, D.~C., \& Abbot, D.~S. 2014, The Astrophysical
  Journal, 787, L2

\bibitem[{{Yoder}(1995)}]{Yoder_1995cnt}
{Yoder}, C.~F. 1995, in Global Earth Physics: A Handbook of Physical Constants
  (American Geophysical Union, Washington D.C), 1--31

\bibitem[{{Zechmeister} {et~al.}(2019){Zechmeister}, {Dreizler}, {Ribas},
  {Reiners}, {Caballero}, {Bauer}, {B{\'e}jar}, {Gonz{\'a}lez-Cuesta},
  {Herrero}, {Lalitha}, {L{\'o}pez-Gonz{\'a}lez}, {Luque}, {Morales},
  {Pall{\'e}}, {Rodr{\'\i}guez}, {Rodr{\'\i}guez L{\'o}pez}, {Tal-Or},
  {Anglada-Escud{\'e}}, {Quirrenbach}, {Amado}, {Abril}, {Aceituno},
  {Aceituno}, {Alonso-Floriano}, {Ammler-von Eiff}, {Antona Jim{\'e}nez},
  {Anwand-Heerwart}, {Arroyo-Torres}, {Azzaro}, {Baroch}, {Barrado},
  {Becerril}, {Ben{\'\i}tez}, {Berdi{\~n}as}, {Bergond}, {Bluhm},
  {Brinkm{\"o}ller}, {del Burgo}, {Calvo Ortega}, {Cano}, {Cardona
  Guill{\'e}n}, {Carro}, {C{\'a}rdenas V{\'a}zquez}, {Casal},
  {Casasayas-Barris}, {Casanova}, {Chaturvedi}, {Cifuentes}, {Claret},
  {Colom{\'e}}, {Cort{\'e}s-Contreras}, {Czesla}, {D{\'\i}ez-Alonso}, {Dorda},
  {Fern{\'a}ndez}, {Fern{\'a}ndez-Mart{\'\i}n}, {Fuhrmeister}, {Fukui},
  {Galad{\'\i}-Enr{\'\i}quez}, {Gallardo Cava}, {Garcia de la Fuente},
  {Garcia-Piquer}, {Garc{\'\i}a Vargas}, {Gesa}, {G{\'o}ngora Rueda},
  {Gonz{\'a}lez-{\'A}lvarez}, {Gonz{\'a}lez Hern{\'a}ndez},
  {Gonz{\'a}lez-Peinado}, {Gr{\"o}zinger}, {Gu{\`a}rdia}, {Guijarro}, {de
  Guindos}, {Hatzes}, {Hauschildt}, {Hedrosa}, {Helmling}, {Henning},
  {Hermelo}, {Hern{\'a}ndez Arabi}, {Hern{\'a}ndez Casta{\~n}o}, {Hern{\'a}ndez
  Otero}, {Hintz}, {Huke}, {Huber}, {Jeffers}, {Johnson}, {de Juan},
  {Kaminski}, {Kemmer}, {Kim}, {Klahr}, {Klein}, {Kl{\"u}ter}, {Klutsch},
  {Kossakowski}, {K{\"u}rster}, {Labarga}, {Lafarga}, {Llamas}, {Lamp{\'o}n},
  {Lara}, {Launhardt}, {L{\'a}zaro}, {Lodieu}, {L{\'o}pez del Fresno},
  {L{\'o}pez-Puertas}, {L{\'o}pez Salas}, {L{\'o}pez-Santiago}, {Mag{\'a}n
  Madinabeitia}, {Mall}, {Mancini}, {Mandel}, {Marfil}, {Mar{\'\i}n Molina},
  {Maroto Fern{\'a}ndez}, {Mart{\'\i}n}, {Mart{\'\i}n-Fern{\'a}ndez},
  {Mart{\'\i}n-Ruiz}, {Marvin}, {Mirabet}, {Monta{\~n}{\'e}s-Rodr{\'\i}guez},
  {Montes}, {Moreno-Raya}, {Nagel}, {Naranjo}, {Narita}, {Nortmann}, {Nowak},
  {Ofir}, {Oshagh}, {Panduro}, {Parviainen}, {Pascual}, {Passegger}, {Pavlov},
  {Pedraz}, {P{\'e}rez-Calpena}, {P{\'e}rez Medialdea}, {Perger}, {Perryman},
  {Rabaza}, {Ram{\'o}n Ballesta}, {Rebolo}, {Redondo}, {Reffert}, {Reinhardt},
  {Rhode}, {Rix}, {Rodler}, {Rodr{\'\i}guez Trinidad}, {Rosich}, {Sadegi},
  {S{\'a}nchez-Blanco}, {S{\'a}nchez Carrasco}, {S{\'a}nchez-L{\'o}pez},
  {Sanz-Forcada}, {Sarkis}, {Sarmiento}, {Sch{\"a}fer}, {Schmitt},
  {Sch{\"o}fer}, {Schweitzer}, {Seifert}, {Shulyak}, {Solano}, {Sota}, {Stahl},
  {Stock}, {Strachan}, {Stuber}, {St{\"u}rmer}, {Su{\'a}rez}, {Tabernero},
  {Tala Pinto}, {Trifonov}, {Veredas}, {Vico Linares}, {Vilardell}, {Wagner},
  {Wolthoff}, {Xu}, {Yan}, \& {Zapatero Osorio}}]{Zechmeister_etal_2019}
{Zechmeister}, M., {Dreizler}, S., {Ribas}, I., {et~al.} 2019, \aap, 627, A49

\end{thebibliography}

\end{document}